\documentclass[aps,amsmath,amssymb,nofootinbib,unsortedaddress]{revtex4} 
\usepackage{graphicx} 
\usepackage{amsmath}
\usepackage{amsfonts,amsbsy}
\usepackage{amssymb}
\usepackage{xcolor}
\usepackage{braket}
\usepackage{cancel}
\usepackage{subcaption}
\usepackage[utf8]{inputenc}

\def\be{\begin{equation}}
\def\ee{\end{equation}}
\def\bea{\begin{eqnarray}}
\def\eea{\end{eqnarray}}

\renewcommand{\v}[1]{ \ensuremath{ {\underline{#1}} }}

\usepackage{tikz}
\usetikzlibrary{calc}

\usepackage[compat=1.1.0]{tikz-feynman}

\newlength{\mysm}
\begin{document}

\title{Incoherent diffractive dijet production and gluon Bose enhancement in the nuclear wave function}

\author{Tiyasa Kar}
\email{tkar@ncsu.edu} 
    \affiliation{Department of Physics, North Carolina State University, Raleigh, NC 27695, USA}

\author{Alexander Kovner}
\email{alexander.kovner@uconn.edu} 
\affiliation{Physics Department, University of Connecticut, 2152 Hillside Road, Storrs, CT 06269, USA}

\author{Ming Li}
\email{li.13449@osu.edu} 
    \affiliation{Department of Physics, The Ohio State University, Columbus, OH 43210 , USA}

\author{Vladimir V. Skokov}
\email{VSkokov@ncsu.edu} 
    \affiliation{Department of Physics, North Carolina State University, Raleigh, NC 27695, USA}
    %\affiliation{RIKEN BNL Research Center, Brookhaven National Laboratory, Upton, NY 11973, USA}

\begin{abstract}
We investigate the effect of gluon Bose enhancement in the nuclear wave function on the dijet production in incoherent diffractive processes in DIS and ultraperipheral collisions.  We demonstrate that Bose enhancement leads to an enhancement of diffractive dijet production cross section when the transverse momenta of the two jets are aligned at zero relative angle. This enhancement is maximal when the magnitude of the transverse momenta of the two jets are equal, and disappears rather quickly as a function of the ratio of the two momenta. We study both the dilute limit and fully nonlinear dense regime where the nuclear wave function is evolved with the leading order JIMWLK equation. In both cases we observe a visible effect, with it being enhanced by the evolution due to the dynamical generation of the color neutralization scale.
\end{abstract}

\maketitle

\section{Introduction}
The investigation of dijet production in Deep Inelastic Scattering (DIS) is one of the key measurements~\cite{Aschenauer:2017jsk} of the upcoming Electron-Ion Collider (EIC), as it provides insights into the phenomenon of gluon saturation at high energies~\cite{Gribov:1984tu,Mueller:1985wy,Mueller:1999wm,Mueller:2001fv,Iancu:2003xm,Gelis:2010nm,Kovchegov:2012mbw}. 

Most of the studies so far, with rare exceptions~\cite{Mantysaari:2019hkq}, have been concerned with the nearly back-to-back dijets~\cite{Marquet:2007vb,Boer:2010zf,Dominguez:2011wm,Dominguez:2011br,Metz:2011wb,Pisano:2013cya,Mueller:2013wwa,Kotko:2015ura,Dumitru:2015gaa,Dumitru:2016jku,vanHameren:2016ftb,Marquet:2016cgx,Kotko:2017oxg,Marquet:2017xwy,Dumitru:2018kuw,Iancu:2018hwa,Klein:2019qfb,Mantysaari:2019hkq,Iancu:2020mos,Hatta:2021jcd,Boussarie:2021ybe,Fujii:2020bkl,Rodriguez-Aguilar:2023ihz}.  
Recently the dijet calculation in DIS in the Color Glass Condensate (CGC) formalism was extended to NLO order in Refs.~\cite{Caucal:2021ent,Caucal:2023nci,Caucal:2023fsf}\, with particular attention to understanding the interplay of high energy evolution and the Sudakov suppression.

In this paper we are interested in a particular aspect of dijet production, namely can it be used as a probe of Bose enhancement of gluons in the nuclear wave function.
This prompts us to consider both inclusive and diffractive dijet production~\cite{Bartels:1999tn,Altinoluk:2015dpi,Hatta:2016dxp,Hagiwara:2017fye,Mantysaari:2019csc,Salazar:2019ncp,Mantysaari:2019hkq,Iancu:2021rup,Hatta:2022lzj,Beuf:2022kyp,Iancu:2022lcw,Rodriguez-Aguilar:2023ihz}; and also to go beyond the  back-to-back kinematics.

In two recent papers \cite{Kovner:2021lty,Kovner:2023yas},  we have studied an observable that allows to directly probe Bose-Einstein correlations between gluons in the nucleus~\cite{Dumitru:2010iy,Altinoluk:2015uaa,Kovner:2018azs}. 
These correlations are very interesting since they provide nontrivial information about two particle correlations in the nuclear wave function~\cite{Gaunt:2009re,Blok:2010ge, Diehl:2011yj, Blok:2013bpa,Diehl:2017wew} which may significantly extend our understanding  beyond the conventional single particle distributions traditionally probed in DIS experiments~\cite{Dumitru:2015gaa, Hatta:2016dxp, Dumitru:2018kuw, Mantysaari:2019hkq}. 
The observable we considered was rather challenging experimentally, as it requires the measurement of three jets and correlations between them. In this work, we argue that gluon Bose-Einstein correlations can be measured in a simpler setup in diffractive dijet production.

Quantum statistics that is responsible for the  universal source of correlations between (identical) gluons is Bose enhancement.  In the hadron wave function this effect leads to the amplification 
in the number of pairs of gluons with the same quantum numbers and small relative momentum, see more in Ref.~\cite{Dumitru:2010iy,Altinoluk:2015uaa,Kovner:2018azs}.

The reason we expect dijet production to be sensitive to the Bose correlation is rather transparent. 
At high energy, in the infinite momentum frame,
the virtual photon fluctuates into a quark-antiquark pair (dipole) with transverse momenta $\v{q}_1$ and  $\v{q}_2$, which scatters on the gluon field of the fast-moving hadron target. 
In the final state this leads to two (quark-anti-quark) jets with the 
transverse momenta $\v{p}_{1}$ and  $\v{p}_{2}$. 
In the hadronic wave function prior to the scattering gluons are correlated via Bose enhancement. That is, the probability of finding two gluons is enhanced if they have the same color and transverse momentum ($\v{k}_{1}  \approx \v{k}_{2}  $). Thus there is an enhanced probability that as a result of scattering, both the quark and the antiquark absorb target gluons with the same transverse momentum  $\v{k}_{1} $.  Since the two gluons are also in the color singlet state, as is the original dipole one can further isolate this contribution by considering diffractive dijet production. The two jets may be away from the back-to-back regime and may even have almost the same transverse momenta (while longitudinal momentum can be very different), if the exchange with the target is due to hard gluons. 

 The dijet observable is much simpler and better experimentally accessible than the trijet we have considered before. An additional advantage is that in the diffractive configuration there is no radiation outside of the dijet. Thus the momentum imbalance of the dijet is not affected by the Sudakov radiation \cite{Sudakov:1954sw,Mueller:2013wwa} and is determined entirely by the momentum exchange with the target.

In this paper we study numerically the diffractive dijet production cross section as a function of the relative angle between the transverse momenta of the two jets in the CGC approach. 
We start with the dilute regime where the calculation can be done to lowest order in the density of the target. We then study the full nonlinear case where the target is evolved by the leading order JIMWLK evolution. 

For perturbative calculation the target averaging is done utilizing the McLerran-Venugopalan (MV) model~\cite{McLerran:1993ka,McLerran:1993ni}. For comparison we also introduce an IR regulator which enforces color neutralization in the target within a finite transverse distance $1/m$. The initial condition for JIMWLK evolution is given by the same MV model. JIMWLK evolution generates the color neutralization scale dynamically. The inverse neutralization scale $m$ grows with rapidity  proportionally to the saturation momentum.

In both cases we find an enhancement of the production for small relative angle. This enhancement is most significant when the transverse momenta of the two jets are equal in magnitude, and disappears when the ratio of the two momenta is of order 1.5-2. The effect is enhanced when the nonvanishing color neutralization scale, and is more pronounced for the JIMWLK evolved results.
 
The paper is structured as follows. In Section II we recap the calculation of the dijet cross section within the CGC approach. In Section III we review the Bose enhancement in the CGC wavefunction as it appeared in calculations of particle production in CGC. In Section IV we calculate the diffractive dijet production in the dilute limit. In Section V we present the results of the full nonlinear, JIMWLK evolved cross section. Finally in Section VI we briefly discuss our results.

\section{Inclusive and incoherent diffractive dijet productions}
\label{Sec:Dijet}
In this section, we review the differential cross section of DIS dijet production within the Color Glass Condensate (CGC) formalism~\cite{Iancu:2003xm,Gelis:2010nm,Kovchegov:2012mbw}. As we alluded to in the introduction, in this framework, the incident photon undergoes a fluctuation into a quark-antiquark pair, which then interacts with the nucleus. We consider the photon to be  right-moving while the nuclear target moves to the left. We set the reference frame such that the photon has a large longitudinal momentum $p^+$ and zero transverse momentum; photon's virtuality is denoted by  $Q^2$. The target nucleus carries a large $P^-$ component, that is, we have  (neglecting the nucleus mass)
\begin{equation*}
    p^\mu=\bigg(p^+,\frac{-Q^2}{2p^+},\v 0\bigg)
\end{equation*}
for the photon
and 
\begin{equation*}
    P^\mu=\bigg(0,P^-,\v 0\bigg)
\end{equation*}
for the nucleus. 
At high energy, it is convenient to work in the so-called dipole picture of DIS, where the photon splits into quark-anti-quark pair with the lifetime of the dipole being much longer than the interaction time with the nucleus. We work in the eikonal approximation, in which the transverse separation between the quark and antiquark remains unchanged while the dipole traverses through the nucleus. Corrections to eikonal approximation are suppressed by powers of collision energy $\sqrt{s}$  \cite{Altinoluk:2022jkk, Chirilli:2018kkw, Cougoulic:2022gbk, Li:2023tlw} and although this suppression is not overwhelming at EIC we will not consider such corrections in the present work.
 
 The expressions for the differential dijet production cross-sections can be found in Ref. ~\cite{Dominguez:2011wm}:
\begin{align}
\label{CS}
    \frac{d\sigma ^{\gamma_{T,L}^{\ast }A\rightarrow q\bar{q}X}}{d^3k_1d^3k_2}=
    &N_{c}\alpha _{em}e_{q}^{2}\delta(p^+-k_1^+-k_2^+)\int\frac{\text{d}^{2}\v x_1}{(2\pi)^{2}}\frac{\text{d}^{2}\v x_1^{\prime }}{(2\pi )^{2}}\frac{\text{d}^{2}\v x_2}{(2\pi)^{2}}\frac{\text{d}^{2}\v x_2^{\prime }}{(2\pi )^{2}}e^{-i\v k_1\cdot(\v x_1-\v x_1^{\prime})} e^{-i\v k_2\cdot (\v x_2-\v x_2^{\prime})}\nonumber\\
    &\times\sum_{\lambda\alpha\beta}\psi^{T,L\lambda}_{\alpha\beta}(\v x_1-\v x_2)\psi^{T,L\lambda*}_{\alpha\beta}(\v x_1^\prime-\v x_2^\prime)\, \mathcal{N}.
    %\left[1+S^{(4)}_{x_g}(x_1,x_2;x_2^{\prime },x_1^{\prime})
%-S^{(2)}_{x_g}(x_1,x_2)-S^{(2)}_{x_g}(x_2^{\prime },x_1^{\prime })\right] 
\end{align}
The wave functions of the transverse and longitudinal photon are
\begin{equation}
\psi^{T\lambda}_{\alpha\beta}(p^+,z,\v r)=2\pi\sqrt{\frac{2}{p^+}}
\begin{cases}
    i\epsilon_fK_1(\epsilon_f|\v r|)\frac{\v r\cdot\v \epsilon^{(1)}}{|\v r|}\left[\delta_{\alpha+}\delta_{\beta+}\bar z+\delta_{\alpha-}\delta_{\beta-}z\right],&\text{if }\lambda=1\\
    i\epsilon_fK_1(\epsilon_f|\v r|)\frac{\v r\cdot\v \epsilon^{(2)}}{|\v r|}\left[\delta_{\alpha-}\delta_{\beta-}\bar z+\delta_{\alpha+}\delta_{\beta+}z\right],&\text{if }\lambda=2
\end{cases}
\end{equation}
 for the transverse ($T$)
and  
\begin{equation}
    \psi^{L\lambda}_{\alpha\beta}(p^+,z,\v r)=2\pi\sqrt{\frac{4}{p^+}}z\bar zQK_0(\epsilon_f|\v r|)\delta_{\alpha\beta}
\end{equation}
for the longitudinal ($L$) polarizations.  The variable $z$ represents the momentum fraction carried by the quark, and $\v r=\v x_1-\v x_2$ represents the transverse separation of the quark-antiquark pair. The helicities of the (massless) quark and antiquark are denoted by $\alpha$ and $\beta$. Additionally, it is convenient to introduce $\epsilon_f^2=z\bar zQ^2$. The function $K_n$ is the modified Bessel function of the second kind of $n^\text{th}$ order. 
In what follows, we will only consider longitudinal polarization of the virtual photon.

In Eq.\eqref{CS}, $\mathcal{N}$ is the factor  expressing the dipole scattering off the nucleus. For inclusive dijet production it is
\begin{equation}
\label{WInclusive}
\mathcal{N}_{\text{inclusive}}=1+S^{(4)}_{x_g}(\v x_1,\v x_2;\v x_2^{\prime },\v x_1^{\prime})-S^{(2)}_{x_g}(\v x_1,\v x_2)-S^{(2)}_{x_g}(\v x_2^{\prime },\v x_1^{\prime }), 
\end{equation}
while for incoherent diffractive dijet production,
\begin{equation}
\mathcal{N}_{\text{incoherent diffractive}}=S^{(2,2)}_{x_g}(\v x_1,\v x_2;\v x_1^\prime,\v x_2^\prime)-S^{(2)}_{x_g}(\v x_1,\v x_2)S^{(2)}_{x_g}(\v x_1^\prime,\v x_2^\prime).
\end{equation}
 The quadrupole and dipole terms arising in ${\cal N}$ of Eq.~\eqref{CS}, respectively, are as follows:
%\begin{eqnarray}
%\frac{d\sigma ^{\gamma_{L}^{\ast }A\rightarrow q\bar{q}X}}{d^3k_1d^3k_2}
%&=&N_{c}\alpha _{em}e_{q}^{2}
%\delta(x_\gamma -1) \frac{z(1-z) \epsilon_f^2}{(p^+)^2} 8 (2\pi )^{2}
%\int
%\frac{\text{d}^{2}x_1}{(2\pi)^{2}}\frac{\text{d}^{2}x_1^{\prime }}{(2\pi )^{2}}
%\frac{\text{d}^{2}x_2}{(2\pi)^{2}}\frac{\text{d}^{2}x_2^{\prime }}{(2\pi )^{2}} \notag \\
%&&\times e^{-ik_{1 }\cdot(x_1-x_1^{\prime })} e^{-ik_{2 }\cdot (x_2-x_2^{\prime })}
% K_0(\epsilon_f |x_1-x_2|)   K_0(\epsilon_f |x_1^\prime-x_2^\prime|)
%\notag \\
%&&\times \left[1+S^{(4)}_{x_g}(x_1,x_2;x_2^{\prime },x_1^{\prime})
%-S^{(2)}_{x_g}(x_1,x_2)-S^{(2)}_{x_g}(x_2^{\prime },x_1^{\prime })\right] \ ,\label{Eq:dis}
%\end{eqnarray}
\begin{align}
	S^{(4)}_{x_g}(\v x_1,\v x_2;\v x_2^{\prime },\v x_1^{\prime}) &= 
\frac{1}{N_c} {\rm Tr} \langle 
V^\dagger(\v{x}_2)  V(\v{x}_1) 
\left[ V^\dagger(\v{x}^\prime_2)  V(\v{x}^\prime_1) \right]^\dagger
\rangle 
\label{Eq:Quadrup_rewrit}\,,
\\
	S^{(2)}_{x_g}(\v x_1,\v x_2) &= 
\frac{1}{N_c} {\rm Tr} \langle 
V^\dagger(\v{x}_2)  V(\v{x}_1) 
\rangle 	\label{Eq:Dipole_rewrit}
\end{align}
where the Wilson lines $V(\v x)$ (and $V^\dagger(\v x)$) encode multiple scattering of the quark (anti-quark) off the gluon field of the nucleus.  
Here $\langle \ldots \rangle$ denotes averaging over the valence target ensemble.  

The additional quantity that appears in the incoherent diffractive cross section is
\begin{equation}
    S^{(2,2)}_{x_g}(\v x_1,\v x_2;\v x_1^\prime,\v x_2^\prime) = 
    \frac{1}{N_c^2} \left\langle 
     {\rm Tr} \left[   
V^\dagger(\v{x}_2)  V(\v{x}_1) \right]
 {\rm Tr}  \left[
V^\dagger(\v{x}^\prime_1)  V(\v{x}^\prime_2) \right]
\right\rangle 	\,.
\end{equation}

\section{Bose enhancement}
\label{Sec:Bose}

\subsection{Soft gluon density matrix and averaging over target configurations}
CGC is an effective field theory based on the separation of the degrees of freedom into valence and soft sectors as quantified by the fraction of the longitudinal momentum of the hadron they carry. The valence charges are treated as static sources of faster soft degrees of freedom. 

The conventional approach in CGC is to first average over the soft sector and then over the valence target configurations -- equations in the previous section were obtained in this precise way. Although this approach is most convenient for practical calculations it is not very illuminating as far as the quantum correlations in the target are concerned. To analyze the latter, one, first, averages over valence sector and only then considers expectations values of operators of interest in the soft sector. 

The CGC wave function~\cite{Kovner:2015hga,Duan:2020jkz} is given by 
\begin{equation}
| \psi \rangle = | s_v \rangle  \otimes   | v \rangle  \,,
\end{equation}
where $| v \rangle $ is the  state vector characterizing the valence degrees of freedom and  $| s_v \rangle$ is the vacuum of the soft fields in the presence of the valence source. 

In the leading perturbative order
\begin{equation}
    | s_v \rangle=\mathcal{C}| 0\rangle, \quad 	{\cal C}=\exp\left\{2 i {\rm tr} \int_{\v{k}} b^i(\v{k}) \phi^a_i(\v{k}) \right\}\,
\end{equation}
where 
\begin{equation}
\phi_i(\v{k})\equiv a_i^+(\v{k})+a_i(-\v{k})\,
\end{equation}
is the soft gluon field operator with $a^{+}_i$, $a_i$ being the gluon creation and annihilation operators.
The background field $b^i_a$ is determined by the valence color charge density $\rho$ via:
\begin{equation}
    b^i_a(\v{k})=g\rho_a(\v{k})\frac{i\v{k}_i}{k^2} + {\cal O}(\rho^2)\,.
	\label{Eq:b}
\end{equation}
In this and the next section, for the sake of simplicity of presentation, we consider the dilute approximation in the target, that is 
the leading contribution of the color charge density. At the leading order in $\rho(\v{k})$, only gluons with the longitudinal  polarization contribute to ${\cal C}$ and $  | s_v \rangle$. 

The valence wave function $|v\rangle$ is customarily modeled in the so-called McLerran-Venugopalan (MV) model as~\cite{McLerran:1993ni,McLerran:1993ka}. The MV model can be formulated in terms of diagonal matrix elements of the valence density matrix in the color charge density basis:
\begin{equation}\label{mv}
  \langle \rho | v \rangle \langle v |  \rho \rangle=   N e^{-\int_{\v{k}}
 \frac{1}{2\mu^2} \rho_a(\v{k})   
  \rho^*_a(\v{k})}\,,
\end{equation}
where $N$ is the normalization factor and the parameter $\mu^2$ determines the average color charge density in the valence wave function.
 
The full hadron density matrix reads 
\begin{equation}
    \hat{\rho}=|v\rangle\otimes  |s_v\rangle \langle s_v| \otimes \langle v| \,.
\end{equation}
Integrating out the valence modes yields the reduced density matrix for the soft sector
\begin{equation}
    \hat{\rho}_r=  {\rm Tr}_\rho \hat\rho\equiv  \int D \rho\,  \langle \rho | \hat{\rho}  |  \rho \rangle =  \int D \rho\,  \langle \rho |v\rangle\,   |s_v\rangle \langle s_v|    \, \langle    v|  \rho \rangle\,.
\end{equation}
In the particle number representation it was derived in Ref.~\cite{Duan:2020jkz}. The nonvanishing matrix elements of $\hat \rho_r$ between states in the momentum space Fock basis are 
%$\hat \rho_r = \prod_q \otimes \hat \rho_r(q)$ element including the normalization is: 
\begin{align}\label{matel}
     &
     \rho_{n,m,\alpha,\beta} \equiv 
     \langle n_c(\v{q}), m_c(-\v{q}) |\hat{\rho}_r(\v{q})|\alpha_c(\v{q}), \beta_c(-\v{q})\rangle 
	 = (1-R) 
	           % \notag \\ &\times
      \frac{
      \left( n+\beta\right)!
  }{\sqrt{n!m!\alpha!\beta!}}\left( \frac {R}{2} \right)^{n+\beta} 
      \delta_{ \left( n+\beta \right), \left(m+\alpha\right)}\,, \\ &R =\left(1+\frac{\v q^2}{2g^2 \mu^2} \right)^{-1} 
\end{align}
with other elements being zero. The matrix elements are explicitly diagonal in color and momentum. The state $|n_c(\v{q}), m_c(-\v{q})\rangle $ denotes $n$ gluons with momentum $\v{q}$ and $m$ gluons with momentum $-\v{q}$.

\subsection{Bose enhancement in the soft density matrix}
Consider the correlator (only longitudinal polarization of gluons is included)
\begin{align}
    D(\v{k
    }, \v{p}) 
     = 
     {\rm Tr} \left( \hat \rho_r  a_b^+(\v{k}) a_c^+(\v{p}) a_b(\v{k}) a_c(\v{p})   \right). 
\end{align}
This measures the correlation between two gluons drawn from the target.  

This can be easily calculated in the MV model.
Consider first
\begin{align}
\langle a_a^+(\v{k}_1) a_b(\v{k}_2) \rangle =  
    {\rm Tr} \left( \hat \rho_r a_a^+(\v{k}_1) a_b(\v{k}_2)   \right) 
     =  (2\pi)^2 \delta^{(2)}(\v{k}_1-\v{k}_2) \delta_{ab}
     \sum_{n,m} n 
     \rho_{n,m,n,m} 
      = (2\pi)^2  \delta^{(2)}(\v{k}_1-\v{k}_2) \delta_{ab} \frac{g^2 \bar\mu^2}{k_1^2}\,.
\end{align}
The sum is computed in Appendix~\ref{app:a}. 

Additionally 
\begin{align}\label{contract}
\langle a_a^+(\v{k}_1) a^+_b(\v{k}_2) \rangle =  
     (2\pi)^2 \delta^{(2)}(\v{k}_1-\v{k}_2) \delta_{ab}
     \sum_{n,m} \sqrt{n+1} \sqrt{m+1}
     \rho_{n,m,n+1,m+1} 
      = (2\pi)^2  \delta^{(2)}(\v{k}_1+\v{k}_2) \delta_{ab} \frac{g^2 \bar\mu^2}{k_1^2}\,
\end{align}
with the same result for $\langle a_a(\v{k}_1) a_b(\v{k}_2) \rangle $. 

In Ref.~\cite{Duan:2021clk}, it was shown that the density matrix $\rho_r$ is Gaussian, that is $\hat \rho_r
 = \exp \left[ - \int d^2 q \, \beta \omega(\v q) c^+(\v{q}) c(\v{q}) \right] 
$ where $c(\v{q})$ is a linear combination of $a^+(\v{q})$ 
and  $a(-\v{q})$. 
Therefore we can use Wick theorem to find the expectation value of $  D(\v{k
    }, \v{p}) $. We should keep in mind that in addition to the standard Wick contraction $\langle a^\dagger a\rangle$, also the contractions $\langle a^\dagger a^\dagger\rangle$ and $\langle a a\rangle$ are nonvanishing and are given by \eqref{contract}. We get 
\begin{align}\notag
  D(\v{k
    }, \v{p})  &=    
    \left\langle  a_b^+(\v{k}) a_c^+(\v{p}) a_b(\v{k}) a_c(\v{p})   \right\rangle 
     \\&= 
     \left\langle  a_b^+(\v{k}) a_b(\v{k}) 
    \right\rangle 
    \left\langle
    a_c^+(\v{p}) a_c(\v{p})   \right\rangle
    + 
    \left\langle  a_b^+(\v{k}) a_c^+(\v{p}) 
    \right\rangle 
    \left\langle
    a_b(\v{k}) a_c(\v{p})   \right\rangle 
     +
     \left\langle  a_b^+(\v{k}) a_c(\v{p}) 
    \right\rangle 
    \left\langle
    a_c^+(\v{p}) a_b(\v{k})   \right\rangle\,.
\end{align}
Here, as we are interested in density-enhanced terms,  we can ignore the order of the operators, that is we can treat the creation annihilation operators as commuting. We thus obtain 
\begin{align}
  D(\v{k
    }, \v{p})  &=  
    \underbrace{\left( S_\perp (N_c^2-1) \frac{g^2 \bar\mu^2}{\v k^2} \right)}_{n(\v k)}
    \underbrace{
    \left( S_\perp (N_c^2-1) \frac{g^2 \bar\mu^2}{\v p^2} \right)}_{n(\v p)}
    \notag \\ &+
    (2\pi)^2  (N_c^2-1) S_\perp
    \left(\frac{g^2 \bar\mu^2} {\v k^2} \right)^2
    \left[  \delta^{(2)}(\v{k}+\v{p}) +  \delta^{(2)}(\v{k}-\v{p}) \right]\,. 
\end{align}
The first term is the ``classical" independent and factorizing contribution, while  the last term represents the non-trivial effect of Bose correlations.
Note that the correlation contains the sum of two delta functions. The first one leads to the enhancement of gluon correlations for back-to-back kinematics. Although this term is important, in physical processes, it is often overpowered by the stronger effect of the transverse momentum conservation. The later term leads to the enhancement of gluons at the same momentum %(for dense target, it will be nearly the same momentum) 
and this is the effect we are interested in. Note that the sensitivity to the Bose enhancement starts from the observables quartic in creation/annihilation   operators or, in other words, from the order of $\bar \mu^4$.

\section{Dilute approximation}
\label{Sec:Dilute}
We now discuss averaging over the target in dilute approximation. The basic objects we have to average are products of Wilson lines.
The Wilson line in the fundamental representation is 
\begin{align}
    V(\v x) = {\cal P} \exp \left( i g \int_{-\infty}^{+\infty} d x^- t^a A_a^+(x^-, x^+=0,\v x) \right) \,. 
\end{align}
In the covariant gauge 
\begin{align}
    \partial^2 A_a^+(x^-, \v x) = g \rho_a(x^-, \v x) \,.
\end{align}
which yields 
\begin{align}
    A_a^+(x^-, \v x) = g \frac{1}{\partial^2} \rho(x^-, \v x)  = 
    -\frac{g}{2\pi} \int d^2 \v y\  \ln(|\v x-\v y| \Lambda)\  \rho(x^-, \v y) 
       = \frac{g}{2\pi} \int \frac{d^2\v k}{(2\pi)^2} \frac{1}{k^2}  \rho(x^-, \v k) \,.
\end{align}
In the MV model, the correlators of $\rho$ are determined by a Gaussian distribution  with
\begin{align}
    \langle \rho^a(x^-, \v x)\rho^b(y^-, \v y) \rangle  = \delta^{ab} \mu^2(x^-) \delta(x^--y^-)
    \delta^{(2)}(\v x-\v y) \,.
\end{align}
For the field correlators this translates into
\begin{align}
    \langle A^a(x^-, \v x)A^b(y^-, \v y) \rangle  
    = \delta^{ab} g^2 \mu^2(x^-) \delta(x^--y^-) L(\v x-\v y)
\end{align}
where 
$$
 L(\v x-\v y) = \frac{g^2}{(2\pi)^2} \int d^2 z    
 \ln(|\v x-\v z| \Lambda)  \ln(|\v z-\v y| \Lambda)\,.
$$

In the dilute limit one expands in the number of gluon exchanges with the target, which necessitates expansion of the Wilson line. As is clear from the previous discussion, the lowest order in which the effect of Bose enhancement is visible is the fourth order in $\mu$, which means that in the expansion of a single Wilson line we need to keep terms up to fourth order in the field
\begin{align}
    V(x) \approx
    1 
    &+  i g \int_{-\infty}^{+\infty} d x^- t^a A_a^+(x^-, \v x)  
    +  (i g)^2 \int_{-\infty}^{+\infty} d x_0^- \int_{-\infty}^{x_0^-} d x_1^-  t^a t^b  A_a^+(x_0^-, \v x)  A_b^+(x_1^-, \v x)  
    \notag \\ &+ 
      (i g)^3 \int_{-\infty}^{+\infty} d x_0^- \int_{-\infty}^{x_0^-} d x_1^-  \int_{-\infty}^{x_1^-} d x_2^-  t^a t^b t^c A_a^+(x_0^-, \v x)  A_b^+(x_1^-, \v x) A_c^+(x_2^-, \v x)  
    \notag \\ &+ 
      (i g)^4 \int_{-\infty}^{+\infty} d x_0^- \int_{-\infty}^{x_0^-} d x_1^-  \int_{-\infty}^{x_1^-} d x_2^-  \int_{-\infty}^{x_2^-} d x_3^-  t^a t^b t^c t^d A_a^+(x_0^-, \v x)  A_b^+(x_1^-, \v x) A_c^+(x_2^-, \v x) A_d^+(x_3^-, \v x) \,. 
\end{align}
To make our notation compact, we introduce the averaged quantities, 
\begin{align}
    \bar \mu^2 = \int_{-\infty}^{+\infty} dx^- \mu^2(x^-)
\end{align}
and 
\begin{align}
   \alpha_a(\v x) =\int_{-\infty}^{+\infty} d x_0^-  A_a^+(x_0^-, \v x)\,.
\end{align}
We also notice that the cubic order in the field can only appear in the expectation values when multiplied by the linear order. We thus can contract two remaining fields already in the expansion of the Wilson line, that is 
\begin{align}
    &\int_{-\infty}^{+\infty} d x_0^- \int_{-\infty}^{x_0^-} d x_1^-  \int_{-\infty}^{x_1^-} d x_2^-  t^a t^b t^c A_a^+(x_0^-, \v x)  A_b^+(x_1^-, \v x) A_c^+(x_2^-, \v x)  \to 
    \int_{-\infty}^{+\infty} d x_0^- \int_{-\infty}^{x_0^-} d x_1^-  \int_{-\infty}^{x_1^-} d x_2^-  t^a t^b t^c 
    \notag \\ & \quad \times \left( 
    A_a^+(x_0^-, \v x)  \langle A_b^+(x_1^-, \v x) A_c^+(x_2^-, \v x) \rangle
    + 
    \langle A_a^+(x_0^-, \v x)  A_b^+(x_1^-, \v x) \rangle A_c^+(x_2^-, \v x) 
        \right. \notag \\ & \left.
        \quad \quad \quad + 
    A_b^+(x_1^-, \v x) \langle A_a^+(x_0^-, \v x)    A_c^+(x_2^-, \v x)  \rangle
    \right) \notag \\  &
    = \frac{C_f L(\v 0)}{2} \int_{-\infty}^{+\infty} d x_0^- \int_{-\infty}^{x_0^-} d x_1^-  t^a A_a^+(x_0^-, \v x)  \mu^2(x_1^-)
    + \frac{C_f L(\v 0)}{2} \int_{-\infty}^{+\infty} d x_0^- \int_{-\infty}^{x_0^-} d x_2^-  t^a A_a^+(x_2^-, \v x)  \mu^2(x_0^-)
     \notag \\ &= \frac{C_f L(\v 0)}{2} \int_{-\infty}^{+\infty} d x_0^- \int_{-\infty}^{+\infty} d x_1^- t^a A_a^+(x_0^-, \v x)  \mu^2(x_1^-) 
     =  \frac{C_f \bar \mu^2 L(\v 0)}{2} t^a \alpha_a(\v x)\,.  
\end{align}
%In this derivation, we took into account that the term  with $A_b^+(x_1^-, \v x) \langle A_a^+(x_0^-, \v x)    A_c^+(x_2^-, \v x)$  vanishes upon integration.. 

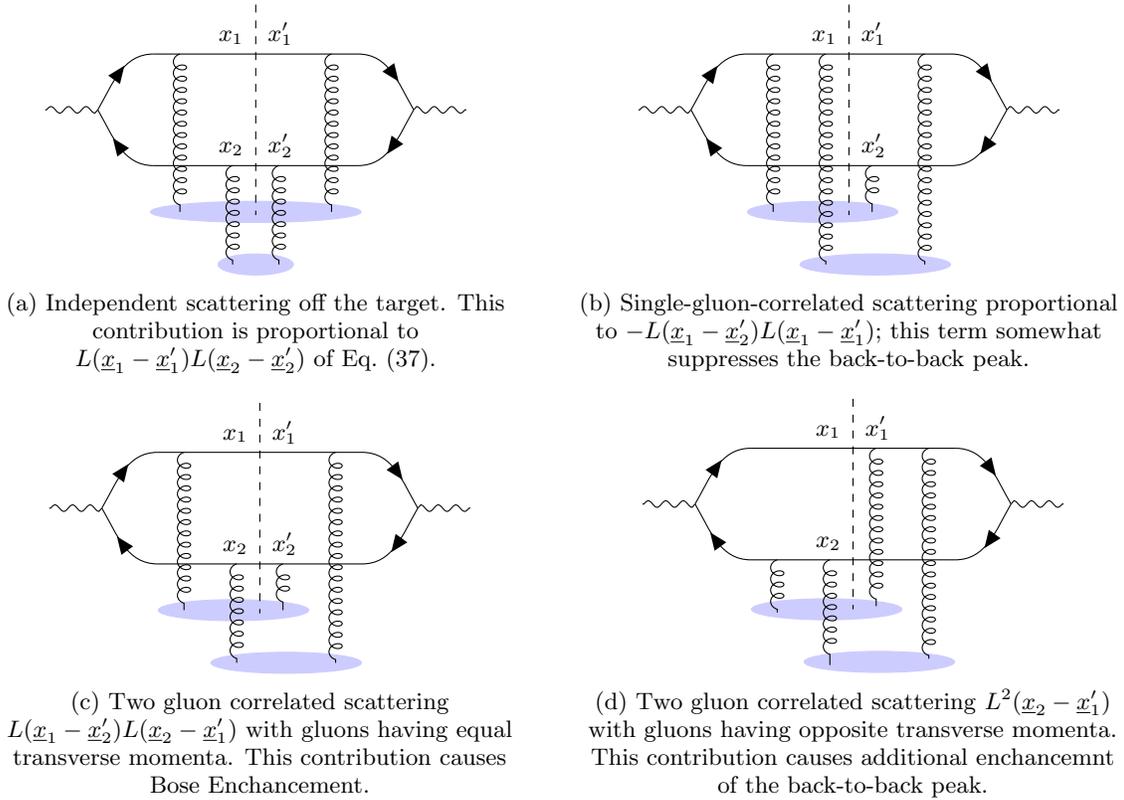
\begin{figure}
\centering
\begin{subfigure}{0.4\textwidth}
\begin{tikzpicture}
  \begin{feynman}
      \vertex (p1);
      \vertex [right=1\mysm of p1] (a1) ; 
      \vertex [above right=1.5\mysm of a1] (g1); % {\(g_1\)};
      \vertex [below right=1.5\mysm of a1] (g2); % {\(g_2\)};
      \vertex [right=6\mysm of a1] (a2) ; %{\(a_2\)};
      \vertex [above left=1.5\mysm of a2] (g3) ; % {\(g_3\)};
      \vertex [below left=1.5\mysm of a2] (g4) ; %{\(g_4\)};
      
      \vertex [right=0.5\mysm of g1] (i1) ; %{\(i_1\)};
      \vertex [left=0.5\mysm of g3] (i2) ; %{\(i_2\)};
      \vertex [right=1.5\mysm of g2] (i3) ; %{\(i_3\)};
      \vertex [left=1.5\mysm of g4] (i4) ; %{\(i_4\)};

      \vertex [above right=2 \mysm and 3\mysm of a1] (f1) ; %{\(f_1\)};
      \vertex [below right=2 \mysm and 3\mysm of a1] (f2) ; %{\(f_2\)};

      \vertex [below=3\mysm of i1] (t1) ; %{\(t_1\)};
      \vertex [below=3\mysm of i2] (t2) ; %{\(t_2\)};
      
      \vertex [below right=1 \mysm and 1\mysm of t1] (t3) ; %{\(t_3\)};
      \vertex [below left=1\mysm and 1\mysm of t2] (t4); % {\(t_4\)};
      \vertex [right=1\mysm  of a2] (p2);
      
       \draw[fill,blue!50, opacity = .4] ( $ (t1)!0.5!(t2) $) ellipse (1.4cm and 0.14cm);
      
       \draw[fill,blue!50, opacity = .4] ( $ (t3)!0.5!(t4) $) ellipse (0.5cm and 0.14cm);
   
    \diagram* {
    {
    (a1) -- [fermion, out=60, in=180] (g1) -- 
    (g1) -- [edge label=\(x_1\quad x_1'\)] (g3) -- [fermion, out=0, in=120] (a2)
    -- [fermion, out=240, in=0] (g4) -- [edge label'=\(x_2\quad x_2'\)]  (g2) -- [fermion, out=180, in =-60 ](a1)
    },
    {
    (f1)--[scalar] (f2)
    },
    {[edges=gluon]
    (i1)-- (t1),
    (i3)-- (t3),
    (i4)-- (t4),
    (i2)-- (t2)
    },
    {[edges=photon]
    (p1)--(a1),
    (p2)--(a2), 
    }
   };
   \end{feynman}
\end{tikzpicture}
    \caption{Independent scattering off the target. This contribution is proportional to $L(\v{x}_1-\v{x}'_1) L(\v{x}_2-\v{x}'_2)$ of Eq.~\eqref{Eq:DiffCorr}.}
    \label{Diag:first}
\end{subfigure}
\hspace{0.5cm}
\begin{subfigure}{0.4\textwidth}
\begin{tikzpicture}
  \begin{feynman}
      \vertex (p1);
      \vertex [right=1\mysm of p1] (a1) ; 
      \vertex [above right=1.5\mysm of a1] (g1); % {\(g_1\)};
      \vertex [below right=1.5\mysm of a1] (g2); % {\(g_2\)};
      \vertex [right=6\mysm of a1] (a2) ; %{\(a_2\)};
      \vertex [above left=1.5\mysm of a2] (g3) ; % {\(g_3\)};
      \vertex [below left=1.5\mysm of a2] (g4) ; %{\(g_4\)};
      
      \vertex [right=0.5\mysm of g1] (i1) ; %{\(i_1\)};
      \vertex [left=0.5\mysm of g3] (i2) ; %{\(i_2\)};
      \vertex [right=1.5\mysm of g1] (i3) ; %{\(i_3\)};
      \vertex [left=1.5\mysm of g4] (i4) ; %{\(i_4\)};

      \vertex [above right=2 \mysm and 3\mysm of a1] (f1) ; %{\(f_1\)};
      \vertex [below right=2 \mysm and 3\mysm of a1] (f2) ; %{\(f_2\)};

      \vertex [below=3\mysm of i1] (t1) ; %{\(t_1\)};
      \vertex [below=4\mysm of i2] (t2) ; %{\(t_2\)};

      \vertex [below right=1 \mysm and 1\mysm of t1] (t3) ; %{\(t_3\)};
      \vertex [above left= 1\mysm and 1\mysm of t2] (t4); % {\(t_4\)};
      \vertex [right=1\mysm  of a2] (p2);
      
       \draw[fill,blue!50, opacity = .4] ( $ (t1)!0.5!(t4) $) ellipse (1.cm and 0.14cm);
      
       \draw[fill,blue!50, opacity = .4] ( $ (t2)!0.5!(t3) $) ellipse (1cm and 0.14cm);
   
    \diagram* {
    {
    (a1) -- [fermion, out=60, in=180] (g1) -- 
    (g1) -- [edge label=\(x_1\quad x_1'\)] (g3) -- [fermion, out=0, in=120] (a2)
    -- [fermion, out=240, in=0] (g4) -- [edge label'=\(\phantom{x_2}\quad x_2'\)]  (g2) -- [fermion, out=180, in =-60 ](a1)
    },
    {
    (f1)--[scalar] (f2)
    },
    {[edges=gluon]
    (i1)-- (t1),
    (i3)-- (t3),
    (i4)-- (t4),
    (i2)-- (t2)
    },
    {[edges=photon]
    (p1)--(a1),
    (p2)--(a2), 
    }
   };
   \end{feynman}
\end{tikzpicture}
    \caption{Single-gluon-correlated scattering proportional to  $-L(\v{x}_1-\v{x}'_2) L(\v{x}_1-\v{x}'_1)$; this term somewhat suppresses the back-to-back peak. }
    \label{Diag:second}
\end{subfigure}
\vspace{0.3\mysm}

\centering
\begin{subfigure}{0.4\textwidth}
\begin{tikzpicture}
  \begin{feynman}
      \vertex (p1);
      \vertex [right=1\mysm of p1] (a1) ; 
      \vertex [above right=1.5\mysm of a1] (g1); % {\(g_1\)};
      \vertex [below right=1.5\mysm of a1] (g2); % {\(g_2\)};
      \vertex [right=6\mysm of a1] (a2) ; %{\(a_2\)};
      \vertex [above left=1.5\mysm of a2] (g3) ; % {\(g_3\)};
      \vertex [below left=1.5\mysm of a2] (g4) ; %{\(g_4\)};
      
      \vertex [right=0.5\mysm of g1] (i1) ; %{\(i_1\)};
      \vertex [left=0.5\mysm of g3] (i2) ; %{\(i_2\)};
      \vertex [right=1.5\mysm of g2] (i3) ; %{\(i_3\)};
      \vertex [left=1.5\mysm of g4] (i4) ; %{\(i_4\)};

      \vertex [above right=2 \mysm and 3\mysm of a1] (f1) ; %{\(f_1\)};
      \vertex [below right=2 \mysm and 3\mysm of a1] (f2) ; %{\(f_2\)};

      \vertex [below=3\mysm of i1] (t1) ; %{\(t_1\)};
      \vertex [below=4\mysm of i2] (t2) ; %{\(t_2\)};
      
      \vertex [below right=1 \mysm and 1\mysm of t1] (t3) ; %{\(t_3\)};
      \vertex [above left= 1\mysm and 1\mysm of t2] (t4); % {\(t_4\)};
      \vertex [right=1\mysm  of a2] (p2);
      
       \draw[fill,blue!50, opacity = .4] ( $ (t1)!0.5!(t4) $) ellipse (1.cm and 0.14cm);
      
       \draw[fill,blue!50, opacity = .4] ( $ (t2)!0.5!(t3) $) ellipse (1cm and 0.14cm);
   
    \diagram* {
    {
    (a1) -- [fermion, out=60, in=180] (g1) -- 
    (g1) -- [edge label=\(x_1\quad x_1'\)] (g3) -- [fermion, out=0, in=120] (a2)
    -- [fermion, out=240, in=0] (g4) -- [edge label'=\(x_2\quad x_2'\)]  (g2) -- [fermion, out=180, in =-60 ](a1)
    },
    {
    (f1)--[scalar] (f2)
    },
    {[edges=gluon]
    (i1)-- (t1),
    (i3)-- (t3),
    (i4)-- (t4),
    (i2)-- (t2)
    },
    {[edges=photon]
    (p1)--(a1),
    (p2)--(a2), 
    }
   };
   \end{feynman}
\end{tikzpicture}
    \caption{Two gluon correlated scattering 
    $L(\v{x}_1-\v{x}'_2) L(\v{x}_2-\v{x}'_1)$ with gluons having equal transverse momenta. This contribution causes Bose Enchancement.  
    }
    \label{Diag:third}
\end{subfigure}
\hspace{0.5cm}
\begin{subfigure}{0.4\textwidth}
\begin{tikzpicture}
  \begin{feynman}
      \vertex (p1);
      \vertex [right=1\mysm of p1] (a1) ; %{\(a_1\)};
      \vertex [above right=1.5\mysm of a1] (g1); % {\(g_1\)};
      \vertex [below right=1.5\mysm of a1] (g2); % {\(g_2\)};
      \vertex [right=6\mysm of a1] (a2) ; %{\(a_2\)};
      \vertex [above left=1.5\mysm of a2] (g3) ; % {\(g_3\)};
      \vertex [below left=1.5\mysm of a2] (g4) ; %{\(g_4\)};
      
      \vertex [right=0.5\mysm of g2] (i1) ; %{\(i_1\)};
      \vertex [left=0.5\mysm of g3] (i2) ; %{\(i_2\)};
      \vertex [right=1.5\mysm of g2] (i3); %{\(i_3\)};
      \vertex [left=1.5\mysm of g3] (i4)  ;%{\(i_4\)};

      \vertex [above right=2 \mysm and 3\mysm of a1] (f1) ; %{\(f_1\)};
      \vertex [below right=2 \mysm and 3\mysm of a1] (f2) ; %{\(f_2\)};

      \vertex [below=1\mysm of i1] (t1) ; %{\(t_1\)};
      \vertex [below=4\mysm of i2] (t2) ; %{\(t_2\)};
      
      \vertex [below=2\mysm of i3] (t3) ; %{\(t_3\)};
      \vertex [below=3\mysm  of i4] (t4); % {\(t_4\)};

      \vertex [right=1\mysm  of a2] (p2);

       \draw[fill,blue!50, opacity = .4] ( $ (t1)!0.5!(t4) $) ellipse (1.cm and 0.14cm);
      
       \draw[fill,blue!50, opacity = .4] ( $ (t2)!0.5!(t3) $) ellipse (1.cm and 0.14cm);
   
    \diagram* {
    {
    (a1) -- [fermion, out=60, in=180] (g1) -- 
    (g1) -- [edge label=\(x_1\quad x_1'\)] (g3) -- [fermion, out=0, in=120] (a2)
    -- [fermion, out=240, in=0] (g4) -- [edge label'=\(x_2\quad \phantom{x_2'}\)]  (g2) -- [fermion, out=180, in =-60 ](a1)
    },
    {
    (f1)--[scalar] (f2)
    },
    {[edges=gluon]
    (i1)-- (t1),
    (i3)-- (t3),
    (i4)-- (t4),
    (i2)-- (t2)
    },
    {[edges=photon]
    (p1)--(a1),
    (p2)--(a2), 
    }
   };
   \end{feynman}
\end{tikzpicture}
    \caption{Two gluon correlated scattering 
    $L^2(\v{x}_2-\v{x}'_1)$ with gluons having opposite transverse momenta. This contribution causes additional enchancemnt of the back-to-back peak.  
    }
    \label{Diag:fourth}
\end{subfigure}

 \caption{Sample of representative diagrams contributing to diffractive dijet production; note that the final state dijet is required to be a color singlet.}
    \label{fig:diags}
\end{figure}

Similar logic allows us to simplify the quartic term (here only one contraction survives due to color structure):  
\begin{align}
    &\int_{-\infty}^{+\infty} d x_0^- \int_{-\infty}^{x_0^-} d x_1^-  \int_{-\infty}^{x_1^-} d x_2^-  \int_{-\infty}^{x_2^-} d x_3^-  t^a t^b t^c t^d A_a^+(x_0^-, \v x)  A_b^+(x_1^-, \v x) A_c^+(x_2^-, \v x) A_d^+(x_3^-, \v x) \nonumber \\ \to 
    &\int_{-\infty}^{+\infty} d x_0^- \int_{-\infty}^{x_0^-} d x_1^-  \int_{-\infty}^{x_1^-} d x_2^-  \int_{-\infty}^{x_2^-} d x_3^-  t^a t^b t^c t^d \langle A_a^+(x_0^-, \v x)  A_b^+(x_1^-, \v x) \rangle \langle A_c^+(x_2^-, \v x) A_d^+(x_3^-, \v x) \rangle 
    \nonumber \\ =  
    & \frac12 \left(\frac{C_f \bar \mu^2 L(\v 0)}{2} \right)^2\,.
\end{align}
We thus effectively have
\begin{align}
     V(\v x) \approx
    & \, 1 + \frac{(ig)^4 }2 \left(\frac{C_f g^2 
 \bar \mu^2 L(\v 0)}{2} \right)^2
    +  i g  t^a \alpha_a(\v x)
    \left(1 + (ig)^2  \frac{C_f \bar g^2  \mu^2 L(\v 0)}{2}  \right)
   \notag \\ & +  (i g)^2 \int_{-\infty}^{+\infty} d x_0^- \int_{-\infty}^{x_0^-} d x_1^-  t^a t^b  A_a^+(x_0^-, \v x)  A_b^+(x_1^-, \v x)\,.  %\notag \\ &
\end{align}
Now, to evaluate the diffractive cross-section, it is also useful to compute the dipole operator to $\mu^4$ order 
\begin{align}
    \frac{1}{N_c}{\rm Tr}\, V^\dagger (\v y)  V (\v x) 
    &= 
    1 + (ig)^4  \left(\frac{C_f g^2 
 \bar \mu^2 L(\v 0)}{2} \right)^2 \notag \\ &
- (ig)^2 \frac{1}{N_c}{\rm Tr} (t_a t_b)  \alpha_a(\v x) \alpha_b(\v y)
 \notag \\ & 
  + (ig)^2   \frac{1}{N_c}{\rm Tr} (t_a t_b) \int_{-\infty}^{+\infty} d y_0^- \int^{-\infty}_{y_0^-} d y_1^-    A_a^+(y_0^-, \v y)  A_b^+(y_1^-, \v y) 
    \notag \\ & 
  + (ig)^2   \frac{1}{N_c}{\rm Tr} (t_a t_b) 
 \int_{-\infty}^{+\infty} d x_0^- \int_{-\infty}^{x_0^-} d x_1^-   A_a^+(x_0^-, \v x)  A_b^+(x_1^-, \v x)
   \notag \\ & 
   - (ig)^4  \frac{1}{N_c}{\rm Tr} (t_a t_b) 
 \left(C_f g^2 
 \bar \mu^2 L(\v 0)  \right) \alpha_a(\v x) 
 \alpha_b(\v y) 
    \notag \\ & 
 + (i g)^3  \frac{1}{N_c}{\rm Tr}\, (t^b t^c t^a)  \alpha_a(\v x)
 \int_{-\infty}^{+\infty} d y_0^- \int^{-\infty}_{y_0^-} d y_1^-    A_b^+(y_0^-, \v y)  A_c^+(y_1^-, \v y) 
   \notag \\ & 
 - (i g)^3  \frac{1}{N_c}{\rm Tr}\, (t^a t^b t^c)  \alpha_a(\v y)
 \int_{-\infty}^{+\infty} d x_0^- \int_{-\infty}^{x_0^-} d x_1^-   A_b^+(x_0^-, \v x)  A_c^+(x_1^-, \v x)
   \notag \\ & 
   + (i g)^4  \frac{1}{N_c}{\rm Tr}\, (t^a t^b t^c t^d)
   \int_{-\infty}^{+\infty} d y_0^- \int^{-\infty}_{y_0^-} d y_1^-    A_a^+(y_0^-, \v y)  A_b^+(y_1^-, \v y) 
      \notag \\ & \quad \quad 
      \quad \quad \quad 
      \quad 
      \times 
    \int_{-\infty}^{+\infty} d x_0^- \int_{-\infty}^{x_0^-} d x_1^-   A_c^+(x_0^-, \v x)  A_d^+(x_1^-, \v x)\,.
\end{align}
The last term can be immediately simplified as we did for a single Wilson line. The cubic term in $A$ involves a trace of three Gell-Mann matrices and, to this order, will necessarily involve one internal contraction leading to zero contribution. We thus can safely drop the cubic terms.  After this simplification, we obtain 
\begin{align}
\label{Eq:DipFinal}
  \frac{1}{N_c}{\rm Tr}\, V^\dagger (\v y)  V (\v x) 
    &= 
    1 
- (ig)^2 \frac{1}{N_c}{\rm Tr} (t_a t_b)  \alpha_a(\v x) \alpha_b(\v y)
 \notag \\ & 
  + (ig)^2   \frac{1}{N_c}{\rm Tr} (t_a t_b) \int_{-\infty}^{+\infty} d y_0^- \int^{-\infty}_{y_0^-} d y_1^-    A_a^+(y_0^-, \v y)  A_b^+(y_1^-, \v y) 
    \notag \\ & 
  + (ig)^2   \frac{1}{N_c}{\rm Tr} (t_a t_b) 
 \int_{-\infty}^{+\infty} d x_0^- \int_{-\infty}^{x_0^-} d x_1^-   A_a^+(x_0^-, \v x)  A_b^+(x_1^-, \v x) \notag \\ &
 + \frac12 \, (ig)^4  \left(C_f g^2 
 \bar \mu^2 L(\v 0) \right)^2 
    - (ig)^4  
 \left(C_f g^2  \bar \mu^2\right)^2
  L(\v 0)  L(\v x-\v y)
    \notag \\ & 
   + \frac{1}{2} (i g)^4  (C_f g^2 \bar \mu^2)^2
   L^2(\v x-\v y) \notag \\ 
   & = 
    1 
  +\frac{(ig)^2}{2}    \frac{1}{N_c}{\rm Tr} (t_a t_b) (\alpha_a(\v y) - \alpha_a(\v x) )
  (\alpha_b(\v y) - \alpha_b(\v x) )
  \notag \\ &
 + \frac{(ig)^4 (C_f g^2 \bar \mu^2)^2}{2}
 [L(\v 0)-  L(\v x-\v y) ]^2  \,. 
\end{align}
The diffractive cross-section envolves  only a bilinear  of the fluctuation of the dipole amplitude, 
$    \frac{1}{N_c}{\rm Tr}\, V^\dagger (\v y)  V (\v x)  - \left \langle \frac{1}{N_c}{\rm Tr}\, V^\dagger (\v y)  V (\v x) 
 \right\rangle 
$. Thus when computing diffractive production we can drop the first and the last term in Eq.~\eqref{Eq:DipFinal}.
For  the leading contribution we get simply 
\begin{align}
\mathcal{N}_{\text{diffractive}} \approx &
\frac{g^4}{16} 
\langle (\alpha_a(\v x_2) - \alpha_a(\v x_1) )
  (\alpha_a(\v x_2) - \alpha_a(\v x_1) )
  (\alpha_b(\v x_2') - \alpha_b(\v x_1') )
  (\alpha_b(\v x_2') - \alpha_b(\v x_1') ) \rangle \notag \\ & -
   \frac{g^4}{16}
\langle (\alpha_a(\v x_2)  - \alpha_a(\v x_1) )
  (\alpha_a(\v x_2) - \alpha_a(\v x_1) ) \rangle
  \langle
  (\alpha_b(\v x_2') - \alpha_b(\v x_1') )
  (\alpha_b(\v x_2') - \alpha_b(\v x_1') )  \rangle
\end{align}
or, evaluating, the averages (see fig.\ref{fig:diags} for diagrammatic representation)
\begin{align}
\label{Eq:DiffCorr}
 \mathcal{N}_{\text{diffractive}}\approx
    \frac{C_f g^8 \bar \mu ^4}{4
   N_c}
   (L(\v x_1-\v x_1')-L(\v x_1
   -\v x_2')-L(\v x_1'-\v x_2
   )+L(\v x_2-\v x_2'))^2\,.
\end{align}
Similar Wilson line structures in the dilute regime have been analyzed in Ref. \cite{Lappi:2015vta}.

The cross-section therefore is
\begin{align}
    E_1E_2
    \frac{d\sigma ^{\gamma_{L}^{\ast }A\rightarrow q\bar{q}X}}{d^3k_1d^3k_2}\bigg|_D 
    % &\alpha _{em}e_{q}^{2}\delta(p^+-k_1^+-k_2^+)
    % \frac{C_fg^8\bar\mu^4S_\perp}{4(2\pi)^6}
    % \int
    % \frac{\text{d}^2\v q}{(2\pi)^2}
    % L(\v q)L(\v k_1+\v k_2-\v q)
    % \bigg[
    % \bigg(
    % \frac{1}{\epsilon_f^2+\v k_2^2}
    % +\frac{1}{\epsilon_f^2+\v k_1^2}
    % \bigg)^2
    % +\frac{1}{\epsilon_f^2+(\v k_1-\v q)^2}\nonumber\\
    % &\times\bigg(
    % \frac{1}{\epsilon_f^2+(\v k_1-\v q)^2}
    % -2\bigg(
    % \frac{1}{\epsilon_f^2+\v k_1^2}
    % +\frac{1}{\epsilon_f^2+\v k_2^2}
    % \bigg)\bigg)
    % +\frac{1}{\epsilon_f^2+(\v k_2-\v q)^2}
    % \bigg(
    % \frac{1}{\epsilon_f^2+(\v k_2-\v q)^2}
    % -2\bigg(
    % \frac{1}{\epsilon_f^2+\v k_1^2}
    % +\frac{1}{\epsilon_f^2+\v k_2^2}
    % \bigg)\bigg)\nonumber\\
    % &+2\frac{1}{\epsilon_f^2+(\v k_1-\v q)^2}
    % \frac{1}{\epsilon_f^2+(\v k_2-\v q)^2}
    % \bigg]\nonumber\\
    =
    &\alpha _{em}e_{q}^{2}Q^2z^2\bar z^2
    \frac{C_fg^8\bar\mu^4S_\perp}{(2\pi)^4}
    \int
    \frac{\text{d}^2\v q}{(2\pi)^2}
     L(\v q)L(\v k_1+\v k_2-\v q)\nonumber\\
    &
    \times\Bigg(
    \frac{1}{\epsilon_f^2+(\v k_1-\v q)^2}
    -\frac{1}{\epsilon_f^2+\v k_1^2}
    +\frac{1}{\epsilon_f^2+(\v k_2-\v q)^2}
    -\frac{1}{\epsilon_f^2+\v k_2^2}
    \Bigg)^2\,.
\end{align}

In order to analyze the angular correlations we consider the correlation function defined as 
\begin{align}
\label{Eq:CorrFunc}
    C(\Delta \phi, |\v{k}_1|, |\v{k}_2|) = \frac{1}{C_{\rm norm}} 
    \int d\phi_1 d\phi_2  \delta(\Delta \phi - \phi_1 + \phi_2) 
    \frac{d\sigma ^{\gamma_{L}^{\ast }A\rightarrow q\bar{q}X}}{d^3k_1d^3k_2}\bigg|_D 
\end{align}
at a constant value of $z$. The momentum-dependent normalization constant $C_{\rm norm}$ is fixed by requiring 
$$
 \int_{-\pi/2}^{\pi/2} d\Delta \phi \, C(\Delta \phi, |\v{k}_1|, |\v{k}_2|) = 1.  
$$

\begin{figure}
\begin{subfigure}{0.99\textwidth}
    \centering
    \includegraphics[width=0.49\linewidth]{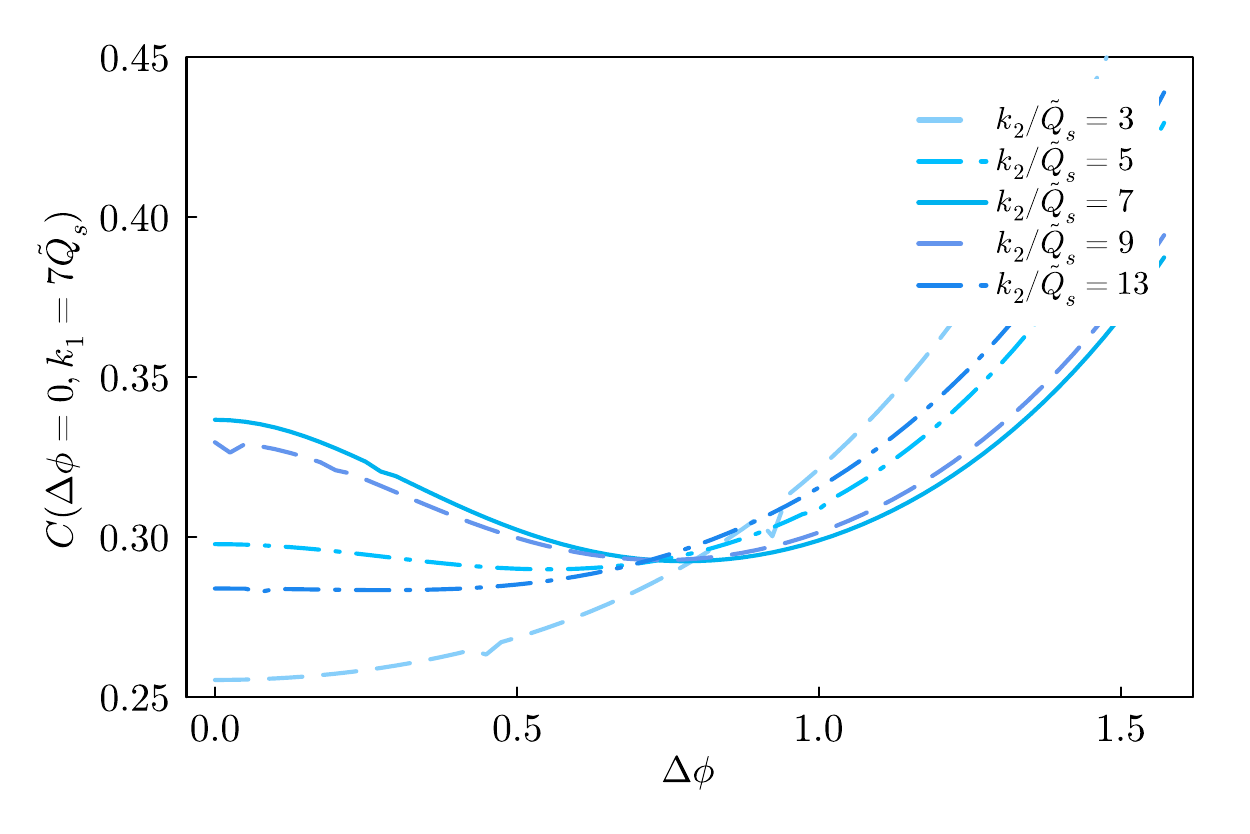}
    \includegraphics[width=0.49\linewidth]{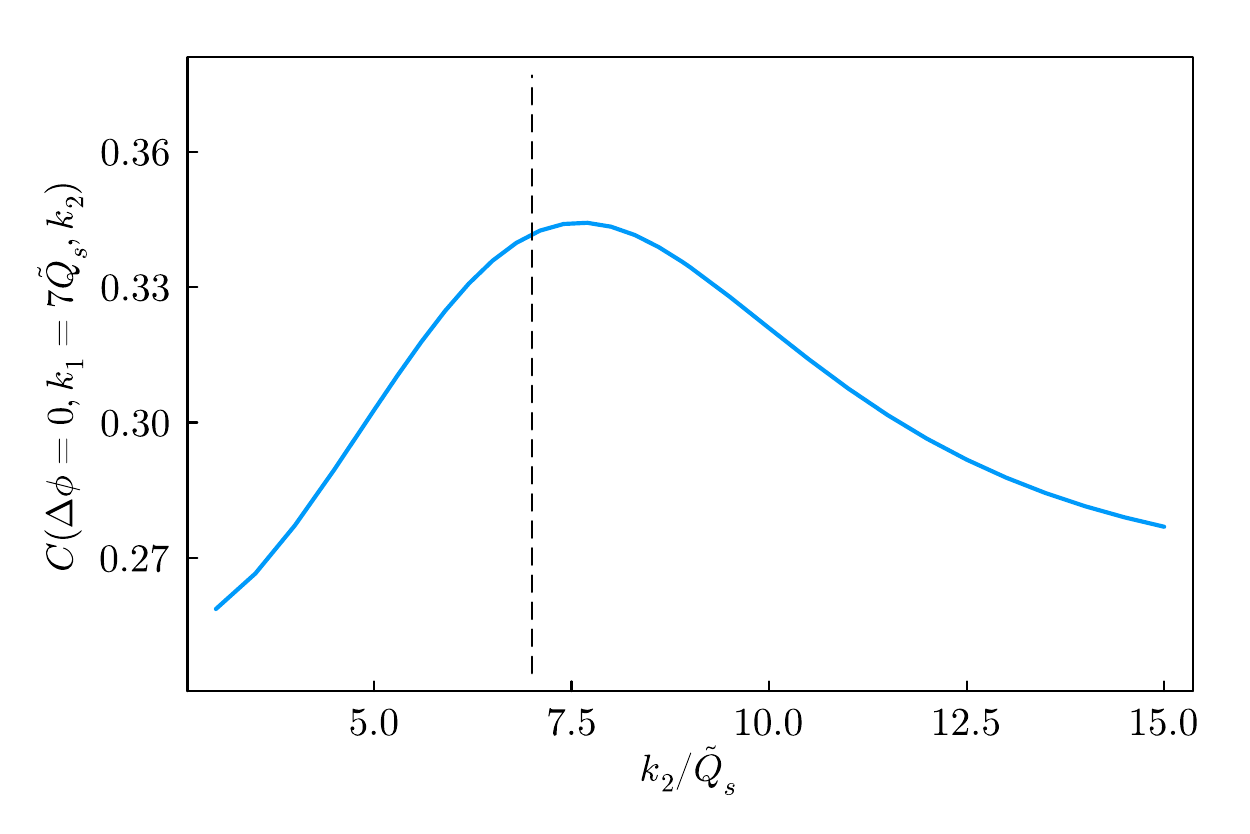}
    \caption{MV model with no color neutralization, $m\to0$.}
            \label{fig:Analytics1} 
\end{subfigure}

\begin{subfigure}{0.99\textwidth}
    \centering
    \includegraphics[width=0.49\linewidth]{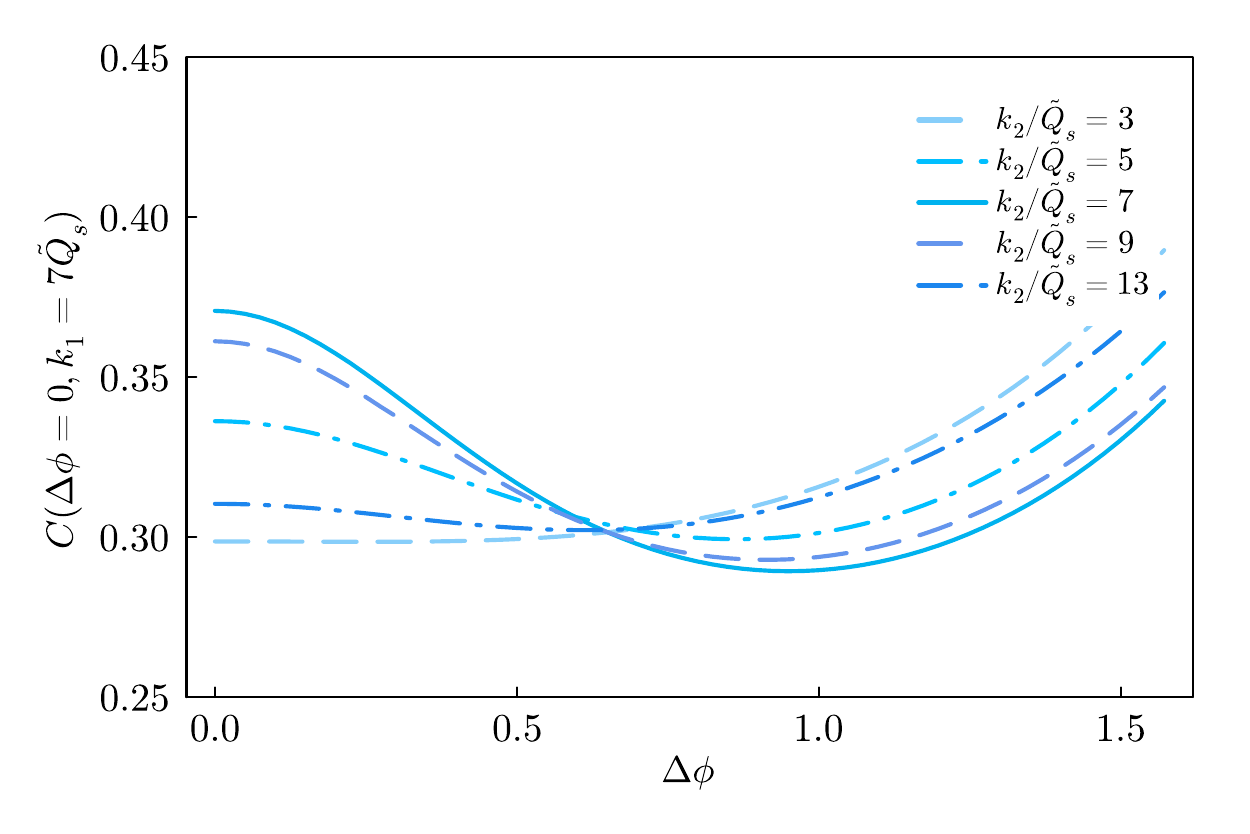}
    \includegraphics[width=0.49\linewidth]{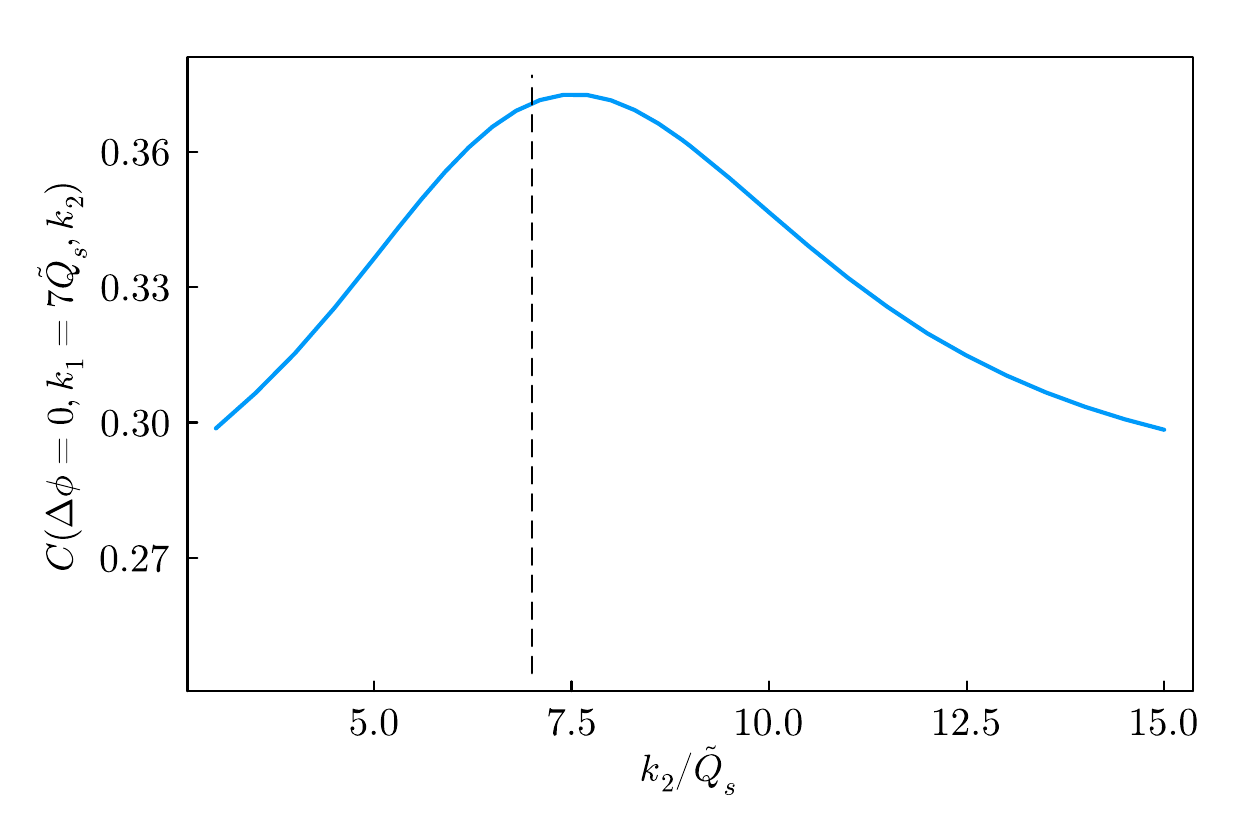}
    \caption{MV model with the color neutralization scale $m=\tilde Q_s$.}
            \label{fig:Analytics2} 
\end{subfigure}
\caption{The correlation function~\eqref{CS} for the incoherent diffractive dijet for $k_1 = 7 \tilde Q_s$ and $\epsilon_f = 2 \tilde Q_s$.   }
        \label{fig:Analytics} 
\end{figure}

The results are plotted in Fig.~\ref{fig:Analytics} as a function of the relative angle between the vectors $\v{k}_1$ and $\v{k}_2$ and their magnitude. The target averaging was performed with two different models: the original MV model, and an MV-like model modified to include a finite color neutralization scale $m$ (see discussion below). The latter is defined by the field-field correlator modified in the IR: 
\begin{align}
\label{eq:IR}
    L(\v q)   = \frac{1}{\v q^2 (\v q^2 + m^2)}\,.
\end{align}

The ``mass''  $m$ plays the role of the color neutralization scale modeled through the color charge correlator (see~Refs.~\cite{Iancu:2002aq,McLerran:2015sva,Kovner:2021lty})
\begin{align}
\label{eq:IRrho}
    \langle \rho^a(x^-, \v{k}) \rho^b(y^-, \v{k}') \rangle
    \propto \frac{k^2}{k^2+m^2} \delta^{ab} \delta(x^- - y^- ) \delta(\v k + \v k') 
\end{align}
which ensures that the correlator  vanishes at momentum $k\ll m$ (or transverse separations greater than $1/m$)  as $\v k^2$ (color neutralization) and that it approaches constant at larger $|\v k|$ (a more realistic model should include anomalous dimension at larger $k$). Equation~\eqref{eq:IR}  follows from 
Eq.~\eqref{eq:IRrho}. 
%It ensures that the color charge density correlator vanishes for transverse separations greater than $1/m$.  %In our illustrative example, we chose $m=0$ and $Q_s$. 
 Although the original MV model does not possess the color  neutralization property, when evolved with the JIMWLK equation to a smaller $x$ a nonvanishing neutralization scale is generated dynamically. 
In this sense, the modified MV model should be viewed as including some of the effects of the low $x$ evolution. Note that we used the model just for the illustration. In the next section, we will use the MV model accompanied by the JIMWLK evolution to compute this observable.   

The figure \ref{fig:Analytics} demonstrates the presence of the peak at the zero relative angle, with the peak strength being the strongest for $|\v{k}_1| \approx |\v{k}_2|$.  We expect that the numerics of the small x evolution will lead to the transition from the pattern of the correlation in Fig.~\ref{fig:Analytics1} to the one in  Fig.~\ref{fig:Analytics2}.

\begin{figure}

    \begin{subfigure}{0.45\textwidth}
        \centering
        \includegraphics[width=\linewidth]{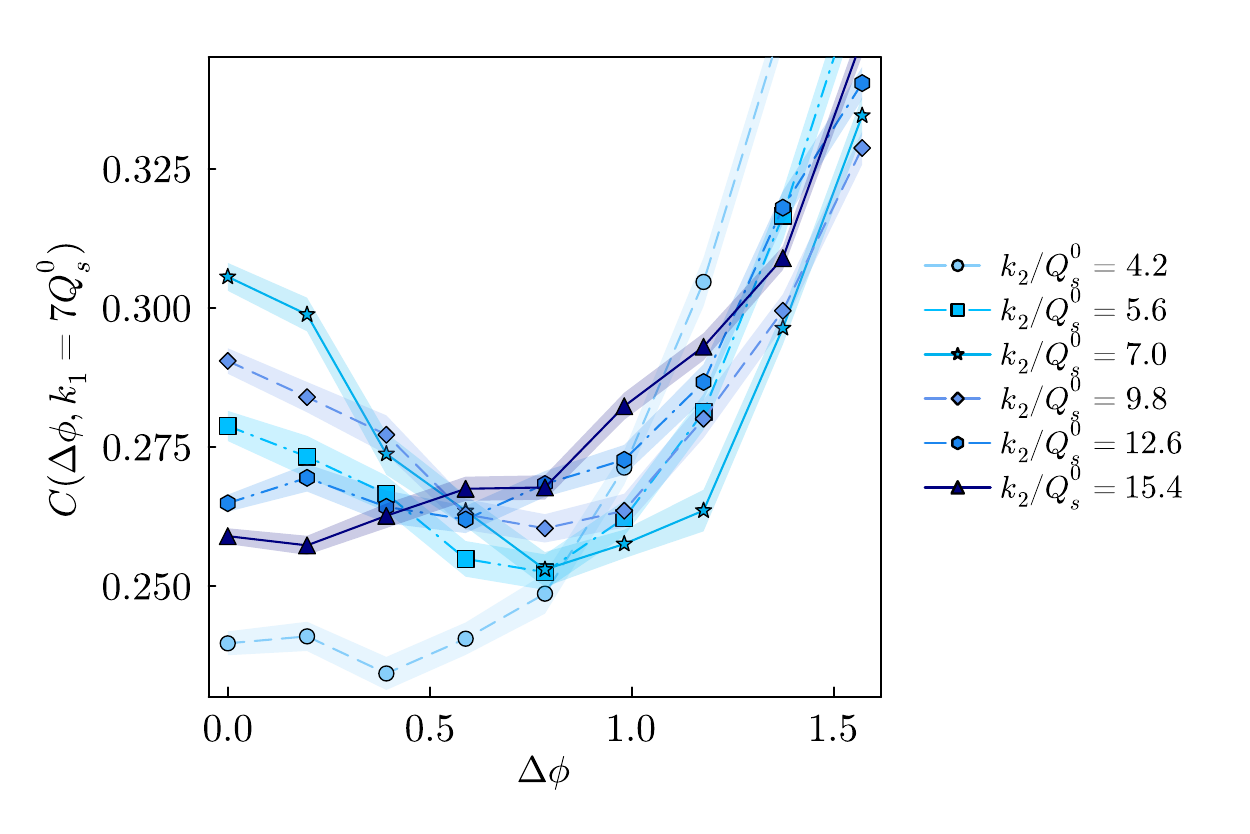}
        \caption{$\alpha_s Y =0.0$}
        \label{fig:0.0}
    \end{subfigure}
    \begin{subfigure}{0.45\textwidth}
        \centering
        \includegraphics[width=\linewidth]{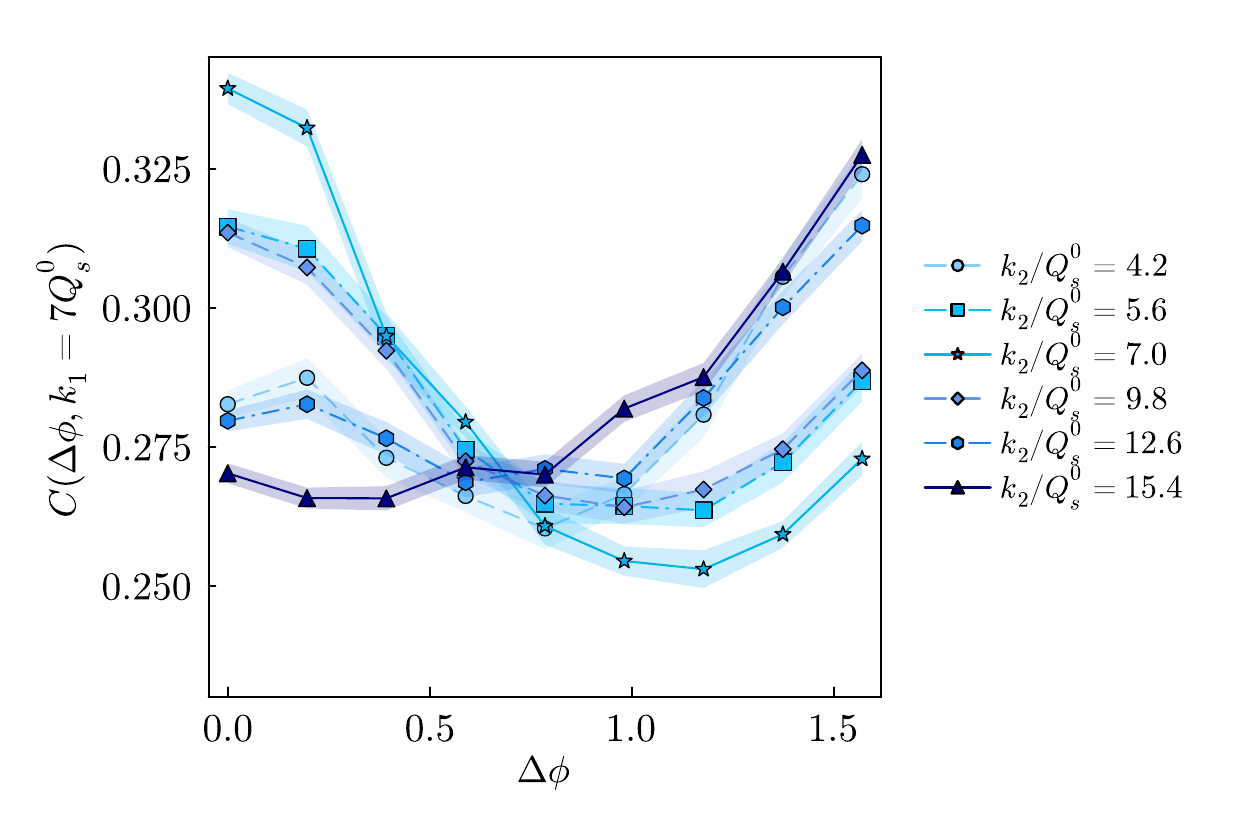}
        
        \caption{$\alpha_s Y =0.4$}
        \label{fig:0.4}
    \end{subfigure}

    \begin{subfigure}{0.45\textwidth}
        \centering
        \includegraphics[width=\linewidth]{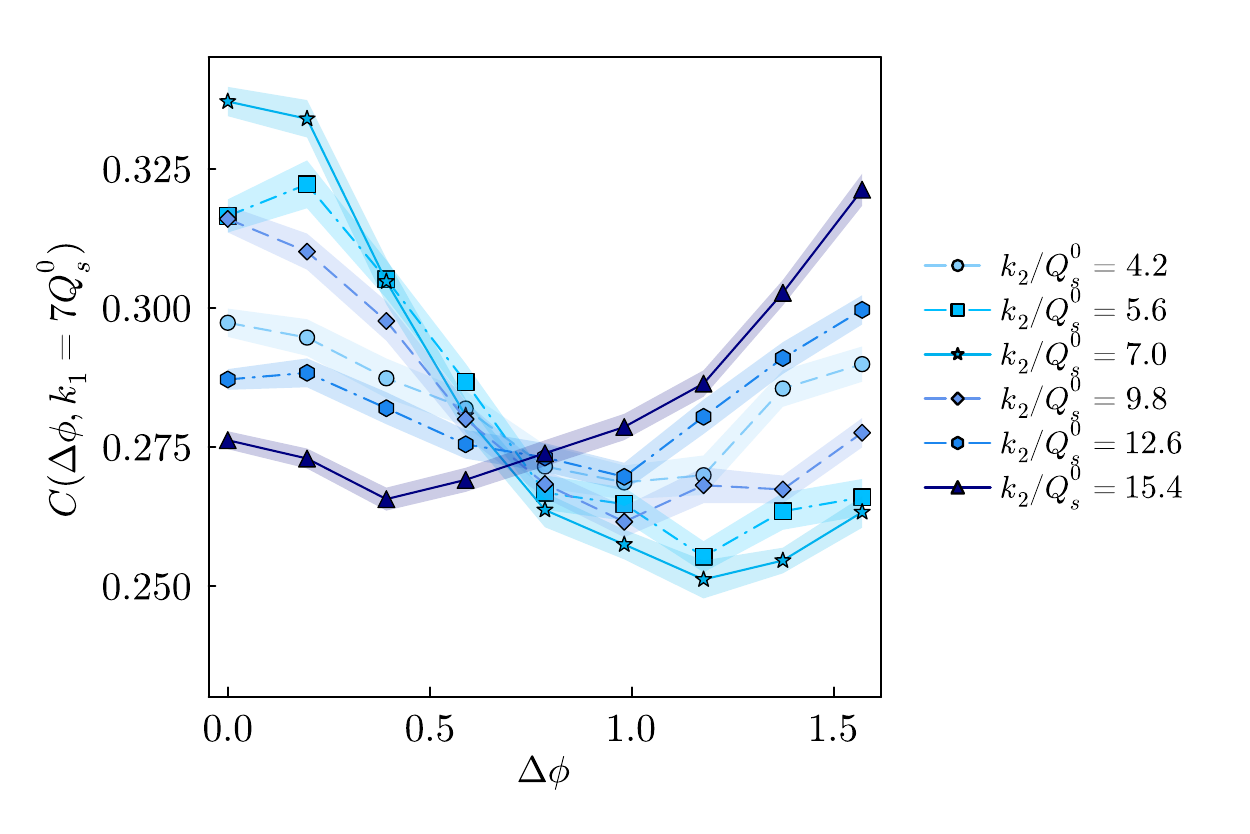}
        \caption{$\alpha_s Y =0.8$}
        \label{fig:0.8}
    \end{subfigure}
    \begin{subfigure}{0.45\textwidth}
        \centering
        \includegraphics[width=\linewidth]{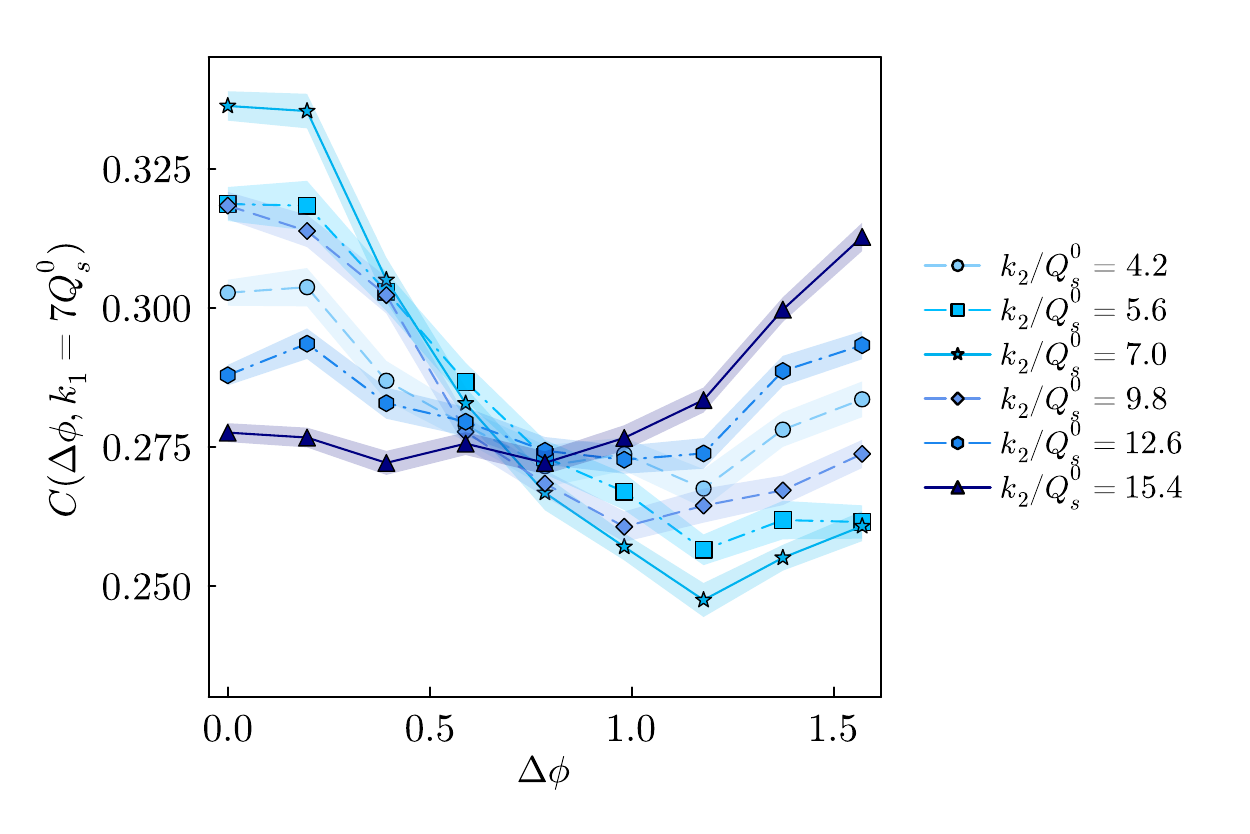}
        \caption{$\alpha_s Y =1.0$}
        \label{fig:1.0}
    \end{subfigure}
        
    \caption{Incoherent Diffractive production as a function of relative angle $k_1=7 Q^0_s$, $\epsilon_f=2 Q^0_s$.  }
    \label{fig:v1}
\end{figure}

Note that the Wilson line correlator entering into the inclusive cross-section to the first nontrivial  order is  
\begin{align}
    \mathcal{N}_{\text{inclusive}} \approx 
    & \frac{g^2}{N_c}\text{Tr}(t^at^b)\bigg(
    \langle
    \alpha_a(\v x_{1})\alpha_b(\v x_{1}^\prime)
    \rangle
    -\langle
    \alpha_a(\v x_{1})\alpha_b(\v x_{2}^\prime)
    \rangle
    -\langle
    \alpha_a(\v x_{2})\alpha_b(\v x_{1}^\prime)
    \rangle
    +\langle
    \alpha_a(\v x_{2})\alpha_b(\v x_{2}^\prime)
    \rangle\bigg)
    %\nonumber\\
    %&+\frac{g^4}{4}\text{Tr}(t^at^bt^ct^d)\bigg(
    %\langle
    % \alpha_a(x_{2\perp}) \alpha_b(x_{2\perp})
    % \alpha_c(x_{1\perp}^\prime)\alpha_d(x_{1\perp}^\prime)
    % \rangle
    % +\langle
    % \alpha_a(x_{2\perp}) \alpha_b(x_{2\perp})
    % \alpha_c(x_{2\perp}^\prime)\alpha_d(x_{2\perp}^\prime)
    % \rangle\nonumber\\
    % &+\langle
    % \alpha_a(x_{1\perp}) \alpha_b(x_{1\perp})
    % \alpha_c(x_{1\perp}^\prime)\alpha_d(x_{1\perp}^\prime)
    % \rangle
    % +\langle
    % \alpha_a(x_{1\perp}) \alpha_b(x_{1\perp})
    % \alpha_c(x_{2\perp}^\prime)\alpha_d(x_{2\perp}^\prime)
    % \rangle\bigg) \nonumber \\ &
    % +g^4C_f\bar\mu^2L(\v 0)\text{Tr}(t^at^b)
    % \bigg(
    % \langle
    % \alpha_a(x_{2\perp})\alpha_b(x_{1\perp}^\prime)
    % \rangle
    % -\langle
    % \alpha_a(x_{2\perp})\alpha_b(x_{2\perp}^\prime)
    % \rangle
    % -\langle
    % \alpha_a(x_{1\perp})\alpha_b(x_{1\perp}^\prime)
    % \rangle
    % +\langle
     %\alpha_a(x_{1\perp})\alpha_b(x_{2\perp}^\prime)
     %\rangle\bigg)\nonumber\\
     %&+g^4\text{Tr}(t^at^bt^ct^d)
     %\langle
%\alpha_a(x_{2\perp})\alpha_b(x_{1\perp})\alpha_c(x_{1\perp}^\prime)\alpha_d(x_{2\perp}^\prime)
     %\rangle
\end{align}
or 
\begin{align}
    \mathcal{N}_{\text{inclusive}} \approx 
    & \frac{C_f g^4 \bar \mu^2}{2}
    \bigg(
    L(\v x_{1} - \v x_{1}^\prime)
    -
    L(\v x_{1} - \v x_{2}^\prime)
     -
    L(\v x_{2} - \v x_{1}^\prime)
    +
    L(\v x_{2} - \v x_{2}^\prime)
    \bigg)\,.
\end{align}
As this expression demonstrates the inclusive production has a non-zero contribution already at order of $\bar \mu^2$. This contribution is not sensitive to quantum gluon correlations. Now at $\bar \mu^4$  order the correlations including Bose Enhancement in the target are present, however they do not result to a peak structure in the angular correlation function at zero angle. The reason why can be understood by considering the diagram in Fig.~\ref{Diag:third}. The diffractive contribution from this diagram is proportional to a quadratic combination of $L$ times $ \delta_{a b'} \delta_{b a'}  \left[\frac{1}{N_c}  \operatorname{tr} (t^a t^b) \right] \left[ \frac{1}{N_c}  \operatorname{tr} (t^{a'} t^{b'}) \right] = \frac{C_f}{2N_c}$, while for the inclusive production we have the same combination of $L$ functions times $ \delta_{a b'} \delta_{b a'}  \left[\frac{1}{N_c}  \operatorname{tr} (t^a t^{a'}   t^{b'} t^b) \right]  = -\frac{C_f}{2N_c}$. 
%Therefore the leading order contribution to the inclusive cross section is insensitive to BE. 
%Upon computing the Wilson line correlator till order of $\bar\mu^4$ we see that there's a dip for $\v k_1\approx\v k_2$. This is because of the negative signs that emerge due to the color algebra.
We see that the contribution of the same Bose enhancement term but with the  sign flipped due to color algebra. 
We thus conclude that  Bose enhancement, which gives an enhancement of the 
zero relative angle peak ($\v k_1\approx\v k_2$) in diffractive production, gives a dip in inclusive! The dip on the background of falling back-to-back peak is impossible to distinguish in a real experiment and this is why we did not pursue its numerical calculation in what follows.

\section{Beyond dilute approximation and small x evolution}
The analysis performed in the previous section relied upon the dilute approximation. In this section, we present the results obtained by numerical simulations of the full MV model with subsequent small-x evolution  computed using leading-order fixed coupling JIMWLK equation.  

JIMWLK is the renormalization group evolution towards smaller x, as conveniently parametrized by $\alpha_s Y = \alpha_s \ln \frac 1 x$. We initialize the evolution with the MV model, generalized to include the color neutralization scale  $m_0=1/4 Q^0_s$ in \eqref{eq:IR}, where $Q_s^0$ is the saturation momentum at zero rapidity. We follow the widely accepted definition of the saturation  scale in the numerical calculation through the fundamental dipole scattering matrix $S(r_s = \sqrt 2 /Q_s) = \exp(-1/2)$. 

As discussed above, the conventional MV model does not incorporate color neutralization. However the JIMWLK evolution generates a color neutralization scale dynamically even if it is not present in the initial condition. This scale turns out to be proportional to the saturation momentum with the proportionality coefficient of order unity, see Ref.~\cite{Duan}. With this in mind we include such a scale in the initial condition as well albeit with a smaller proportionality factor.

Numerical procedures for MV and JIMWLK are  identical to those described in  Ref.~\cite{Lappi:2007ku} and Ref.~\cite{Lappi:2012vw} correspondingly.  
As in the previous section, we fixed the value of $z$ (although it can be chosen arbitrary), in order to compute the correlations functions defined in Eq.~\eqref{Eq:CorrFunc}. The details of the numerical calculation of the diffractive observable are presented in Appendix~\ref{app:b}.

\begin{figure}
    \centering
    \includegraphics[width=0.49\linewidth]{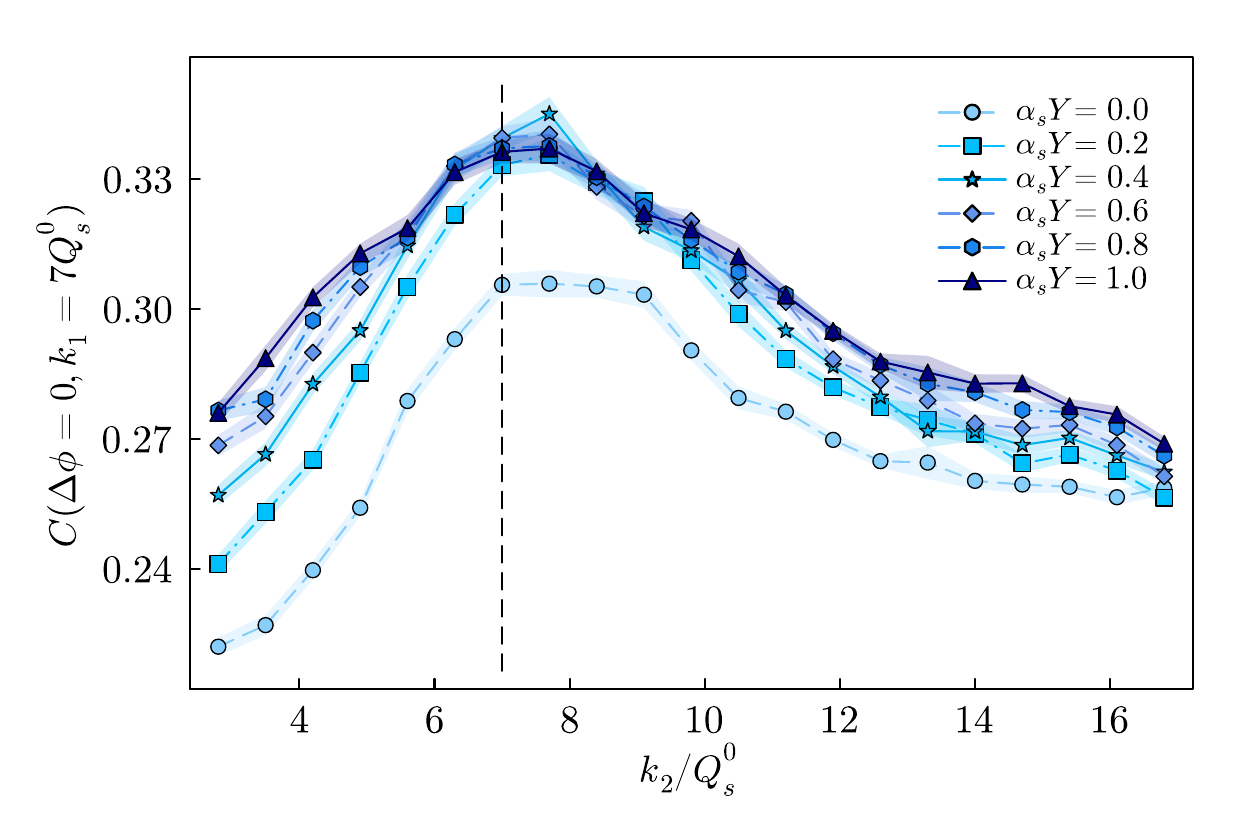}
    \caption{Incoherent Diffractive production for collinear configuration $\Delta \phi=0$ as a function o  f $k_2$: $k_1=7 Q^0_s$, $\epsilon_f=2 Q^0_s$.}
    \label{fig:v}
\end{figure}

The results are presented in the plots, see Figs. \ref{fig:v1} and \ref{fig:v}.

\section{Conclusions}
Let us discuss our numerical results. 

First consider Fig.~\ref{fig:Analytics}. The correlator plotted exhibits the features expected from the Bose enhancement. Both, for $m=0$ and $m=Q_s$ the correlator has a maximum at zero angle as long as the momenta of the two gluons are close to each other. The maximum is most pronounced when $k_1=k_2$ and disappears when the ratio of the two momenta becomes roughly 1.5. Introduction of the color neutralization scale makes the maximum more robust, both in terms of it being more pronounced and surviving for wider range of momenta. This feature was observed also in our earlier work on trijet production \cite{Kovner:2021lty,Kovner:2023yas}. Qualitatively, the origin of this behavior is that finite color neutralization scale suppresses gluons with small transverse momenta. These gluons mostly contribute to the correlator at $\Delta\phi=\pi$, and suppressing them makes the peak at $\Delta\phi=0$ more pronounced. Thus the presence of a large saturation (or color neutralization) scale is a welcome feature from the perspective of observing the Bose enhancement effects.

Moving on to Fig.~\ref{fig:v1} we see the confirmation of the same trend. Evolution in energy generates color neutralization scale \cite{Duan}. This scale quickly grows beyond $Q_s^0$ present in the initial condition. As a result evolving from $\alpha_sY=0$ to $\alpha_sY=.4$ significantly increases the Bose enhancement signal. This is clearly seen moving from Fig.~\ref{fig:0.0} to Fig.~\ref{fig:0.4}.
Interestingly, further evolution to $\alpha_sY=.8$ and beyond changes the picture very little. Although there is marginal enhancement of the signal, it is much less noticeable than the initial change starting with the MV model. 
This suggests that there is importance not only to the value of the saturation scale, but also to the shape of the gluon distribution in the hadron. 

To summarize, we have shown that one can probe Bose enhancement with a simpler observable than discussed earlier - the diffractive dijet production. To do that one needs to go away from the correlation limit into regime where the total momentum and the momentum imbalance of the two jets are close.

\acknowledgements 

A.K. is supported by the NSF Nuclear Theory grant 2208387. This material is based
upon work supported by the U.S. Department of Energy, Office of Science, Office of Nuclear
Physics through the Contract No. DE-SC0020081 (V.S.) and the Saturated Glue (SURGE) Topical Collaboration. 
A.K. and V.S. thank the Binational Science Foundation grant \#2021789 for support. 

V.S. thanks the ExtreMe Matter Institute for partial support and A. Andronic for hospitality at Physics Department of  Muenster University.

\appendix

\section{Useful sums}
\label{app:a}
Using the identity 
\begin{align}
    (n+m)! = \int_0^\infty st 
    e^{-t} t^{n+m}
\end{align}
we get 
\begin{align}
 \label{Eq:Sum1}
\notag 
\sum_{n,m} n 
     \rho_{n,m,n,m} &= 
     (1-R)
     \sum_{n,m}
     n 
     \frac{(n+m)!}{n! m!}
     \left(\frac{R}{2}\right)^{n+m}
    =
     (1-R) \int_0^\infty dt e^{-t}
    \sum_n 
    n \frac{1}{n!} \left(\frac{R t}{2}\right)^{n}
    \sum_m 
 \frac{1}{m!} \left(\frac{R t}{2}\right)^{m}
 \\& = \frac{  (1-R) R}{2}\int_0^\infty dt\, t\,  e^{-t(1-R)} 
 = \frac{R}{2(R-1)} = \frac{g^2\mu^2}{q^2}\,.
\end{align}
The sum
\begin{align}
\notag 
\sum_{n,m} \sqrt{m+1} \sqrt{n+1}
     \rho_{n,m,n+1,m+1}
     &= 
     (1-R)
     \sum_{n,m}
     \sqrt{m+1} \sqrt{n+1}
     \frac{(n+m+1)!}{\sqrt{n! m! (n+1)! (m+1)!}}
     \left(\frac{R}{2}\right)^{n+m+1}
     \\ &= 
     (1-R)
     \sum_{n,m}
     \frac{(n+m+1)!}{n! m! }
     \left(\frac{R}{2}\right)^{n+m+1}
     =     (1-R)
     \sum_{n,m}
     n \frac{(n+m)!}{n! m! }
     \left(\frac{R}{2}\right)^{n+m}
\end{align}
reduces to Eq.~\eqref{Eq:Sum1}.

\section{Numerical evaluation of the incoherent diffractive cross-section}
\label{app:b}
The main difficulty in computing the expectation value for both diffractive and inclusive cross section  is in computing the integrals of  the term involving four Wilson lines  
\begin{align}
T = & \int
\frac{\text{d}^{2}\v x_1}{(2\pi)^{2}}\frac{\text{d}^{2}\v x_1^{\prime }}{(2\pi )^{2}}
\frac{\text{d}^{2}\v x_2}{(2\pi)^{2}}\frac{\text{d}^{2}\v x_2^{\prime }}{(2\pi )^{2}}e^{-i\v k_{1 }\cdot(\v x_1-\v x_1^{\prime })} e^{-i\v k_{2 }\cdot (\v x_2-\v x_2^{\prime })}
 K_0(\epsilon_f |\v x_1-\v x_2|)   K_0(\epsilon_f |\v x_1^\prime-\v x_2^\prime|)
\notag \\
&\times 
\left[V^\dagger(\v{x}_2)  V(\v{x}_1) \right]_{ij}
\left[V^\dagger(\v{x}^\prime_1) V(\v{x}^\prime_2)   \right]_{kl}\,.
\end{align}
To proceed define the amplitude 
\begin{align}
M_{ij}(\v k_1,\v k_2) = & \int
\frac{\text{d}^{2}\v x_1}{(2\pi)^{2}}
\frac{\text{d}^{2}\v x_2}{(2\pi )^{2}}
%\frac{\text{d}^{2}x_2{(2\pi)^{2}}
%\frac{\text{d}^{2}x_2^{\prime }}{(2\pi)^{2}}
e^{-i\v k_{1 }\cdot \v x_1 - i \v k_2 \v x_2} 
%e^{-ik_{2 }\cdot (x_2-x_2^{\prime })}
K_0(\epsilon_f |\v x_1-\v x_2|)   
%K_0(\epsilon_f |x_1^\prime-x_2^\prime|)
\left[V^\dagger(\v{x}_2)  V(\v{x}_1) \right]_{ij}
%\\  & = 
%\frac{1}{2 \pi } 
%\int \frac{\text{d}^{2} p}{(2\pi)^{2}}
%\frac{1}{(p+k_1)^2 + \epsilon_f^2 }
%[
%V^{\dagger}(k_2+k_1+p)
%V(p)
%]_{ij}
%\left[V^\dagger(\vec{x}^\prime_1) V(\vec{x}^\prime_2)   \right]_{kl}
\end{align}
which can be efficiently evaluated using the fast Fourier transform. 

 Once the amplitude is known, we can find the compination  
\begin{align}
    T = M_{ij}(\v k_1,\v k_2) M_{kl}(-\v k_2,-\v k_1)
\end{align}
and thus for the diffractive we simply  
\begin{align}
    T = {\rm tr} M(\v k_1,\v k_2)\,    
    {\rm tr}  M(-\v k_2,-\v k_1)
\end{align}
which requires storing a four dimensional matrix ${\rm tr} M(\v k_1,\v k_2)$ as enumerates by the momentum components. For inclusive one has to store full $M_{ij}(\v k_1,\v k_2)$.  That is computing the inclusive cross section requires a factor of $N_c^2-1$ more operating memory and thus is more challenging.  

Finally, the diffractive cross section boils down to evaluating the averages over the target proportional to 
\begin{align}
    \langle {\rm tr} M(\v k_1,\v k_2)\,    
    {\rm tr}  M(-\v k_2,-\v k_1) \rangle - 
    \langle {\rm tr} M(\v k_1,\v k_2)\,    
    \rangle 
     \langle
    {\rm tr}  M(-\v k_2,-\v k_1) \rangle \,.
\end{align}

\bibliography{notes}

\end{document}